\documentclass[aps, preprint,eshowkeys,nofootinbib]{revtex4-1}
\usepackage{verbatim,graphics,graphicx,color,slashed,textcomp,bbm,mathdots,multirow,array}
\usepackage[normalem]{ulem} 
\usepackage[colorlinks=true,linkcolor=red,urlcolor=blue,citecolor=blue]{hyperref}

\usepackage{autobreak}
\usepackage{float}

\graphicspath{{fig/}}

\makeatletter

\newcommand{\Rmnum}[1]{\expandafter\@slowromancap\romannumeral #1@}
\makeatother

\usepackage[colorlinks=true,linkcolor=red,urlcolor=blue,citecolor=blue]{hyperref}

\begin{document}

\title{Probing Stochastic Ultralight Dark Matter with Space-based Gravitational-Wave Interferometers}
\author{Yue-Hui Yao$^{a,c}$}
\author{Yong Tang$^{a,b,c}$}
\affiliation{\begin{footnotesize}
		${}^a$University of Chinese Academy of Sciences (UCAS), Beijing 100049, China\\
		${}^b$School of Fundamental Physics and Mathematical Sciences, \\
		Hangzhou Institute for Advanced Study, UCAS, Hangzhou 310024, China \\
        ${}^c$International Center for Theoretical Physics Asia-Pacific, Beijing/Hangzhou, China 
		\end{footnotesize}}

\begin{abstract}
    Ultralight particles are theoretically well-motivated dark matter candidates. In the vicinity of the solar system, these ultralight particles can be described as a superposition of plane waves, resulting in a stochastic field with sizable amplitude fluctuations on scales determined by the velocity dispersion of dark matter.
    In this work, we systematically investigate the sensitivity of space-based gravitational-wave interferometers to the stochastic ultralight dark matter field within the frequentist framework. We derive the projected sensitivity of a single detector using the time-delay interferometry. Our results show that space-based gravitational-wave interferometers have the potential to probe unconstrained regions in parameter space and improve the current limit on coupling strengths. 
    Furthermore, we explore the sensitivity of a detector network and investigate the optimal configuration for ultralight dark matter detection. We introduce the overlap reduction function for ultralight dark matter, which quantifies the degree of correlation between the signals observed by different detectors. We find that the configuration, where the signals observed by two detectors are uncorrelated, is the optimal choice for ultralight dark matter detection due to a smaller chance of missing signal. This contrasts with the detection of stochastic gravitational-wave background, where the correlated configuration is preferred. Our results may provide useful insights for potential joint observations involving space-based gravitational-wave detectors like LISA and Taiji, as well as other ULDM detection networks operating in the coherence limit.
\end{abstract}
 
\maketitle

\section{Introduction}

According to the standard model of cosmology $\Lambda$CDM, dark matter~(DM) makes up the dominant part of matter in the universe, but its particle nature remains mystery. 
Ultralight bosonic particles, like axion~\cite{PhysRevLett.38.1440, Weinberg1978, PhysRevLett.40.279, PRESKILL1983127, Abbott:1982af,
Dine:1982ah}, dilaton~\cite{Damour:1994ya,Damour:1994zq,Capozziello:2011et} and dark photon~\cite{PhysRevD.84.103501, PhysRevD.93.103520, Ema:2019yrd, Ahmed:2020fhc, Kolb:2020fwh},
are viable candidates of DM, which not only have strong theoretical motivation but also can mitigate tensions of $\Lambda$CDM on small scales~\cite{deBlok:2009sp,Boylan-Kolchin:2011qkt,Bullock:2017xww,Tulin:2017ara}. 
These ultralight particles typically have mass below $10$~eV~\cite{PhysRevLett.85.1158,PhysRevD.95.043541,Hui:2021tkt}, which implies that their occupation number in a de Broglie volume is numerous if they contribute a non-negligible fraction of the local DM. Therefore, quantum fluctuations are suppressed and these ultralight particles can be  effectively described by classical waves.
These waves with random initial phases travel at different velocities and interfere with each other, resulting in a stochastic field whose amplitude evolves randomly on the time scale~(coherence time) and spatial scale~(coherence length) set by the mass and velocity dispersion of DM~\cite{PhysRevA.97.042506,PhysRevD.97.123006,PhysRevD.103.076018}.

Numerous experiments have been conducted or proposed, spanning different mass bands and exploring diverse phenomenological aspects of ultralight dark matter~(ULDM)~\cite{PhysRevLett.115.011802,PhysRevLett.116.031102,PhysRevLett.117.061301,PhysRevLett.123.031304,PhysRevD.94.015019,PhysRevD.97.055039,XIA2021115470,PhysRevLett.130.181001,PhysRevLett.126.181102,PhysRevLett.126.071302,PhysRevD.107.043004,PhysRevD.83.044026,fukuyama2023axionsupermassiveblackholes,amaral2024vectorwavedarkmatter,PhysRevLett.123.021102,PhysRevLett.124.061102,PhysRevD.107.044053,Kim:2023pkx,Khmelnitsky_2014,Wu:2023dnp,PhysRevD.106.066006,PhysRevD.108.123017,PhysRevD.109.055017,Luu_2024,PhysRevLett.120.241104,PhysRevLett.114.161301,PhysRevLett.115.201301,PhysRevA.93.063630,PhysRevD.98.064051}. 
For instance, if ULDM couples with the Standard Model~(SM) within the low-energy effective action, it could mediate an additional force that is potentially detectable in the fifth-force experiments~\cite{doi:10.1146/annurev.nucl.53.041002.110503}. 
Besides, this force is generally composition-dependent~\cite{PhysRevD.82.084033}, i.e.~its strength varies among different materials, and may induce violations of equivalence principle~(EP), which can be probed by EP tests such as the Lunar Laser Ranging~\cite{PhysRevLett.93.261101,doi:10.1142/S021827180901500X}, experiments conducted by the E\"{o}t-Wash group~\cite{PhysRevLett.100.041101,Wagner_2012}, and the MICROSCOPE mission~\cite{PhysRevLett.119.231101,PhysRevLett.129.121102}.  

ULDM can also imprint observable signatures in gravitational wave~(GW) interferometers by exerting position-dependent oscillatory force on test masses~\cite{ Arvanitaki:2016fyj, PhysRevLett.121.061102, Grote:2019uvn, Vermeulen:2021epa}. 
Constraints from both ground-based GW detectors like LIGO and Virgo~\cite{LIGOScientific:2016aoc} and spacecraft LISA-Pathfinder~\cite{PhysRevLett.120.061101} have been obtained~\cite{Guo:2019ker,fukusumi2023upper, PhysRevD.107.063015}, and projected sensitivities of space-based GW detectors such as LISA~\cite{amaroseoane2017laser}, Taiji~\cite{Hu:2017mde} and TianQin~\cite{Luo_2016} have also been derived~\cite{PhysRevLett.121.061102, PhysRevD.100.123512, PhysRevD.108.083007, PhysRevD.110.023025}, which shows the remarkable ability of GW detectors on exploring unknown regions in parameter space~\footnote{DM can also be probed by GW detectors indirectly through GWs produced by DM-related processes~\cite{10.21468/SciPostPhysCore.3.2.007,PhysRevLett.131.091401,bhattacharya2024continuousgravitationalwavesnew}. Here, we focus on the signals imprinted directly by DM on the detectors.}. 
In the future,  GW detectors are expected to improve current constraints by several times for dilaton-like scalar fields, and several orders of magnitude more if the vector field couples with the baryon number or baryon minus lepton number.
In addition, GW detectors will probe a different direction in the space of coupling parameters, complementing EP tests. 

In most of previous works, the ULDM field is usually treated as deterministic and its stochastic nature is neglected in deriving the sensitivity of detector for simplicity. However, it has been pointed out that accounting for the stochastic nature is crucial for interpreting experiment results and deriving appropriate constraints~\cite{Centers:2019dyn,PhysRevD.104.055037}. 

In this work, we investigate the sensitivity of space-based GW interferometers to the stochastic ULDM field systematically within the frequentist framework, accounting for its stochastic nature through a general likelihood approach. 
While we take a scalar field as an example, our method can be readily extended to vector and tensor fields.
We first derive the sensitivity of a single detector using time-delay interferometry~(TDI)~\cite{Armstrong_1999,PhysRevD.62.042002,Tinto:2020fcc}. 
Considering the potential overlap in duty cycles of multiple missions,
we also investigate the sensitivity of a detector network in typical configurations using the general likelihood formalism. We introduce the overlap reduction function for ULDM,
which measures the correlation between the ULDM signals in different detectors, and discuss how it varies with the separation and relative orientation of detectors.
We find that the non-correlated configuration, where detectors observe independent ULDM signals, outperforms the correlated configuration, since detectors in the non-correlated configuration will probe independent patches of the stochastic field and reduce the probability of missing signal. 
This highlights the advantage of the non-correlated configuration for ULDM detection, contrasting with the detection of stochastic gravitational wave background~(SGWB)~\cite{PhysRevD.59.102001,Romano:2016dpx} where the co-aligned and co-located configuration is favored for its ability to enhance the signal-to-noise ratio through signal correlation.

This paper is organized as follows. In section \ref{Theoretical Framework}, we derive and summarize some useful results of ULDM. Then in section \ref{Detect ULDM}, we specify the interactions between ULDM and the SM and derive the power spectral density~(PSD) of the ULDM signal in TDI channels. Later in section \ref{Single detector} and \ref{Network}, we derive the sensitivity of a single detector and a detector network with the general likelihood formalism, respectively. Here We also make a comparison between ULDM detection and SGWB detection. Finally, we conclude in \ref{Conclusions}.

In this paper, we use natural units ($c = \hbar =1$) unless explicitly stated otherwise.

\section{Theoretical Framework for Ultra-Light Dark Matter} \label{Theoretical Framework}
In this section, we present the theoretical framework of ULDM and clarify notations for our later discussions. Similar treatments can be found in Refs.~\cite{PhysRevA.97.042506,PhysRevD.97.123006,PhysRevD.105.035029,Hui_2021,Hui:2021tkt,nakatsuka2022stochastic}. While the equations in this section are presented in complex form for notational simplicity, we have implicitly taken their real parts before performing any non-linear operations.

\subsection{The local ULDM field}
The local ULDM filed $\Phi(t, \boldsymbol{x})$ can be described as a superposition of monochromatic plane waves 
\begin{equation} \label{plane-wave} 
    \Phi(x) = \sum_{\boldsymbol{v}} \frac{\sqrt{2\rho/N}}{m} \; e^{i\left(\omega t - \boldsymbol{k}\cdot\boldsymbol{x} + \theta_{\boldsymbol{v}}\right)},
\end{equation}
where $\omega = m(1+v^2/2)$, $\boldsymbol{k} = m\boldsymbol{v}$, $m$ is the mass of ULDM particles, \(\theta_{\boldsymbol{v}}\)s are random phases drawn independently from the $[0,2\pi)$ uniform distribution, 
$\rho$~\footnote{We adopt $\rho \approx 0.3~\text{GeV}/\text{cm}^3$ in this work.} is the local energy density of DM, and $N$ denotes the total number of ULDM particles in the vicinity. 

We decompose the sum over velocity in Eq.~(\ref{plane-wave}) into two steps,
\begin{equation}
    \Phi(x) = \sum_{\boldsymbol{v}_i}\sum_{\boldsymbol{v}\in[\boldsymbol{v}_i,\boldsymbol{v}_i+\Delta\boldsymbol{v})} \frac{\sqrt{2\rho/N}}{m} \; e^{i\left(\omega t - \boldsymbol{k}\cdot\boldsymbol{x} + \theta_{\boldsymbol{v}}\right)}.
\end{equation}
We first sum up the components in a small velocity interval $\Delta \boldsymbol{v}$, 
which is small enough that we can ignore the difference between their velocities,
\begin{equation}
    \sum_{\boldsymbol{v}\in[\boldsymbol{v}_i,\boldsymbol{v}_i+\Delta\boldsymbol{v})} \frac{\sqrt{2\rho/N}}{m} \; e^{i\left(\omega t - \boldsymbol{k}\cdot\boldsymbol{x} + \theta_{\boldsymbol{v}}\right)} 
    \simeq \frac{\sqrt{2\rho/N}}{m} e^{i\left[m(1+v_i^2/2) t - \boldsymbol{k}_i\cdot\boldsymbol{x}\right]} \sum_{\boldsymbol{v}\in[\boldsymbol{v}_i,\boldsymbol{v}_i+\Delta\boldsymbol{v})} e^{i\theta_{\boldsymbol{v}}}.
\end{equation}
The sum over random phases is equivalent to a random walk in the complex plane and can be written as
\begin{equation}
    \sum_{\boldsymbol{v}\in[\boldsymbol{v}_j,\boldsymbol{v}_j+\Delta\boldsymbol{v})} e^{i\theta_{\boldsymbol{v}}} = \beta_j e^{i\theta_j},
\end{equation}
where \(\theta_j\) is again a random phase distributed uniformly in $[0,2\pi)$, and \(\beta_j\) a random variable governed by the Rayleigh distribution
\begin{equation} \label{pdf of beta unscaled}
    P\left(\beta_j\right) = \frac{2\beta_j}{N_j}e^{-\beta^2_j/N_j}.
\end{equation}
$N_j$ is the number of terms in the sum $\sum_{\boldsymbol{v}\in[\boldsymbol{v}_i,\boldsymbol{v}_i+\Delta \boldsymbol{v})}$ or equivalently the number of ULDM particles in the velocity interval \([\boldsymbol{v}_j, \boldsymbol{v}_j+\Delta\boldsymbol{v})\) given by 
\begin{equation} \label{N_j=N Delta}
    N_j = N\cdot\Delta_j,\quad\Delta_j\equiv d^3v\;f(\boldsymbol{v}_j),
\end{equation}
with $f(\boldsymbol{v})$ the normalized velocity distribution of DM.

\begin{figure}
    \centering
    \includegraphics[width=\linewidth]{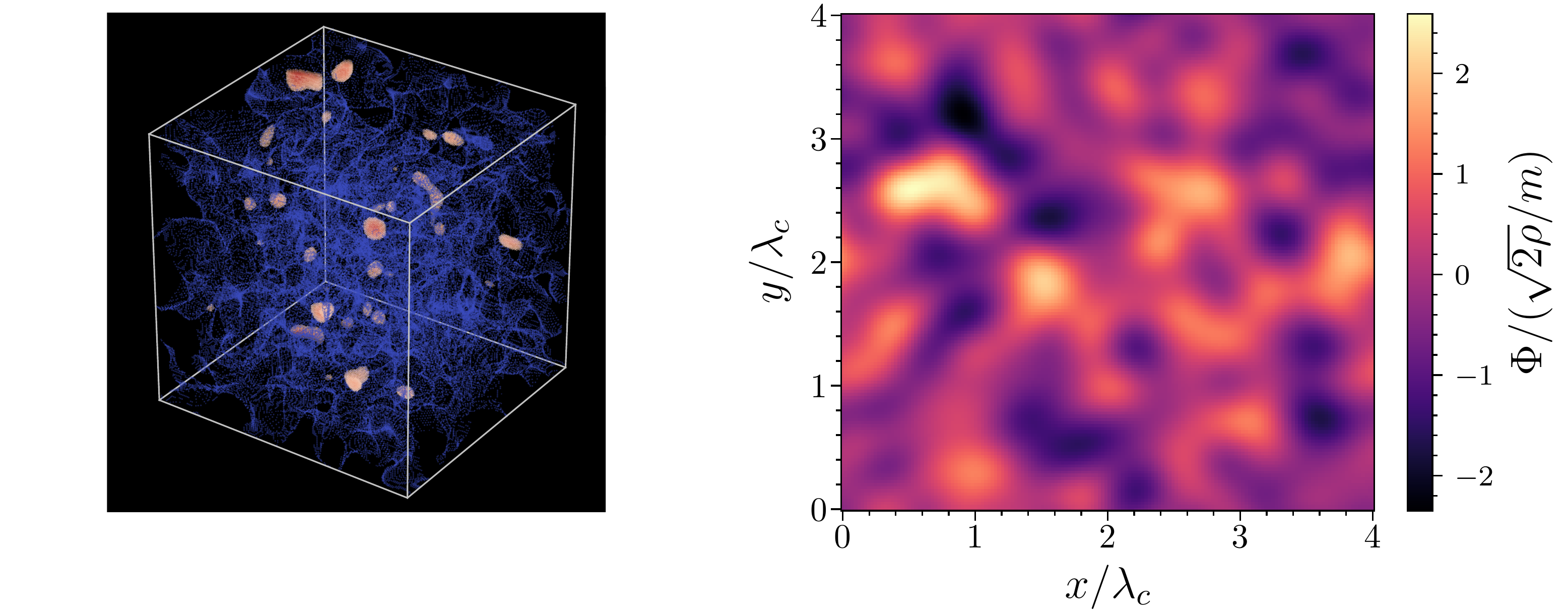}
    \caption{A realization of ULDM field $\Phi(t=\text{const}, \boldsymbol{x})$. 
    (Left) The spatial distribution of density $\rho=\Phi^2$, with high-density regions in red and low-density regions in blue.
    (Right) A 2D slice of $\Phi(x)$ with constant $t$ and $z$. The patterns, arising from interference between waves, have a typical size $\lambda_c^3$. When $|\boldsymbol{x}_1-\boldsymbol{x}_2|\gg \lambda_c$,  the field values $\Phi(t,\boldsymbol{x}_1)$ and $\Phi(t,\boldsymbol{x}_2)$ are uncorrelated. 
    }
    \label{fig:Phi}
\end{figure}

Gathering the above results and rescaling $\beta _j$ by $\sqrt{N_j/2}$ to get rid of the $N_j$ dependence in Eq.~(\ref{pdf of beta unscaled}), Eq.~(\ref{plane-wave}) can be simplified to the form
\begin{equation} \label{Phi rescaled}
    \Phi(x) = \frac{\sqrt{\rho}}{m} \sum_{\boldsymbol{v}_j}  \sqrt{\frac{N_j}{N}} \; \beta_j 
    e^{i\left(\omega_j t - \boldsymbol{k}_j\cdot\boldsymbol{x} + \theta_j\right)}
    = \frac{\sqrt{\rho}}{m} \sum_{\boldsymbol{v}_j}  \sqrt{\Delta_j} \; \beta_j 
    e^{i\left(\omega_j t - \boldsymbol{k}_j\cdot\boldsymbol{x} + \theta_j\right)},
\end{equation}
or using $n,l$ to label the magnitude and direction of velocity, $\boldsymbol{v}=v_n\hat{e}_l$~($|\hat{e}_l|=1$), 
\begin{equation} \label{Phi disc}
    \Phi(x) = \frac{\sqrt{\rho}}{m} \sum_{n}\sum_{l} \sqrt{\Delta_{nl}} \; \beta_{nl} e^{i\left(\omega_{nl}t - \boldsymbol{k}_{nl}\cdot\boldsymbol{x} + \theta_{nl}\right)}.
\end{equation}
Figure.~\ref{fig:Phi} is a 3D snapshot of $\Phi(x)$ generated using Eq.~(\ref{Phi disc}).
As illustrated, $\Phi(x)$ fluctuates stochastically on spatial scale $\lambda_c$ (the coherence length of ULDM field defined in Eq.~(\ref{lc})) and the field values at two points separated by a distance greater than $\lambda_c$ are uncorrelated.

The spatial derivative of $\Phi(x)$ can be readily obtained by differentiating Eq.~(\ref{Phi disc}),
\begin{equation} \label{dPhi disc}
    \nabla\Phi(x) = -i\sqrt{\rho} \sum_{n}\sum_{l} \sqrt{\Delta_{nl}} v_n\beta_{nl} e^{i\left(\omega_{nl}t - \boldsymbol{k}_{nl}\cdot\boldsymbol{x} + \theta_{nl}\right)}\;\hat{e}_l.
\end{equation}

\subsection{Two-point correlation function} \label{2p correlation function}
ULDM field has a stochastic nature, and its statistical properties are encoded entirely in the correlation function.
We calculate the two-point correlation function for ULDM from Eq.~(\ref{Phi disc}) directly~(see Appendix~\ref{appendix 2p func} for details)
\begin{equation} \label{2pp}
    R_{\Phi}(\tau, \boldsymbol{d}) := \langle\Phi(x)\Phi(x')\rangle
	= \frac{\rho}{m^2} \int d^3v\;f(\boldsymbol{v})\;\cos\left[m\left(1+\frac{v^2}{2}\right)\tau - m\boldsymbol{v}\cdot\boldsymbol{d}\right],
\end{equation}
where \(\tau \equiv t'-t\), \(\boldsymbol{d}\equiv\boldsymbol{x}^{\prime}-\boldsymbol{x}\), and $\langle\cdots\rangle$ denotes ensemble average. 
It is seen that
\begin{equation} \label{rho=phi2}
    \left\langle\Phi^2(x)\right\rangle = \frac{\rho}{m^2} \int d^3v f(\boldsymbol{v}) = \frac{\rho}{m^2}\;,  
\end{equation}
where we use the normalization condition of velocity distribution.
Equation.~(\ref{rho=phi2}) indicates that $\Phi$ constitutes the entire local DM.

Correlation functions between the field and its spatial derivative
$R_{i}(\tau, \boldsymbol{d}) \equiv \langle\Phi(x)\partial'_i\Phi(x')\rangle$
and between spatial derivatives themselves
\(R_{ij}(\tau, \boldsymbol{d}) \equiv \langle\partial_i\Phi(x)\partial'_j\Phi(x')\rangle \)
can be derived from Eqs.~(\ref{Phi disc}, \ref{dPhi disc}) through straightforward but tedious calculations. However, there is a shortcut to derive them. 
Note that
$R_{ij}(\tau, \boldsymbol{d}) \equiv \langle\partial_i\Phi(x)\partial'_j\Phi(x')\rangle 
= \partial_i\partial'_j\langle\Phi(x)\Phi(x')\rangle 
= -\partial^{\boldsymbol{d}}_i\partial^{\boldsymbol{d}}_j R_{\Phi}(\tau, \boldsymbol{d})$
with the superscript $\boldsymbol{d}$ indicating that the derivative acts on $\boldsymbol{d}$.
Combining with Eq.~(\ref{2pp}), we have
\begin{align} 
    \label{2pp field+gradient}R_{i}\left(\tau, \boldsymbol{d}\right) &= \frac{\rho}{m} \int d^3v\;f(\boldsymbol{v})v_i\; \sin\left[m\left(1+\frac{v^2}{2}\right)\tau - m\boldsymbol{v}\cdot\boldsymbol{d}\right], \\
    \label{2pp gradient}R_{ij}\left(\tau, \boldsymbol{d}\right) &= \rho \int d^3v\;f(\boldsymbol{v})v_iv_j \;\cos\left[m\left(1+\frac{v^2}{2}\right)\tau - m\boldsymbol{v}\cdot\boldsymbol{d}\right].
\end{align}
Note that while the values of $\left\langle\Phi\partial_i\Phi\right\rangle$ and $\left\langle\partial_i\Phi\partial_j\Phi\right\rangle$, which transforms as a vector and tensor under coordinate transformations, respectively, depend on the choice of the coordinate system, 
the observables are related to quantities like $\hat{n}_i\left\langle\Phi\partial_i\Phi\right\rangle$ and $\hat{n}_i\left\langle\partial_i\Phi\partial_j\Phi\right\rangle\hat{n}_j$, which are scalars under coordinate transformations.
Therefore, we can derive the DM signal in any convenient coordinate system.

When $\tau=\boldsymbol{d}=0$, the integral in Eq.~(\ref{2pp gradient}) reduces to the second moment of the velocity distribution, and Eq.~(\ref{2pp field+gradient}) yields $R_{i}=0$.  
Therefore, for an arbitrary velocity distribution, the field and its gradient at the same spacetime point are uncorrelated.


\subsection{Power spectral density of a ULDM field}
Due to the stochastic nature of a ULDM field, the signal induced by its oscillation is also stochastic, with its properties characterized statistically by the power spectral density~(PSD). As we will see, the signal PSD is related to the PSD of ULDM through the detector's response function.
The two-sided PSD of ULDM is defined as the Fourier transform of the correlation function,
\begin{equation} \label{psd 2pp}
    \begin{split}
    S_{\Phi}(f,\boldsymbol{d}) &:= \int^{\infty}_{-\infty}d\tau  \; e^{-i2\pi f\tau} \; R_{\Phi}(\tau, \boldsymbol{d}) \\
    &= \frac{\rho}{m^2} \int^{\infty}_{-\infty} d\tau  \; e^{-i2\pi f\tau} 
    \times \int dvv^2 \int d\hat{\Omega} \; f(\boldsymbol{v}) \; \cos\left[m\left(1+\frac{v^2}{2}\right)\tau - m\boldsymbol{v}\cdot\boldsymbol{d}\right] \\
    &= \frac{\pi \rho}{m^3}v_f \int d\hat{\Omega} \; f(v_f\hat{\Omega}) \;e^{-imv_f\hat{\Omega}\cdot\boldsymbol{d}},
\end{split}
\end{equation}
where $v_f = \sqrt{2(f/f_c-1)}$, $f_c=m/(2\pi)$, and we neglect the negative frequency part.
From Eq.~(\ref{psd 2pp}), we observe that $S_{\Phi}(f,\boldsymbol{d})$ has a structure closely related to the phase space structure of DM.

If the velocity distribution is isotropic, i.e.\(~f(\boldsymbol{v})=f(v)\), then the directional integral in Eq.~(\ref{psd 2pp}) can be performed analytically and we have
\begin{equation} \label{psd 2ppiso}
        S^{\text{iso}}_{\Phi}(f,\boldsymbol{d}) 
        =  \frac{\pi \rho}{m^3} f(v_f)v_f \int d\hat{\Omega} \; e^{-imv_f\hat{\Omega}\cdot\boldsymbol{d}}
        = \frac{4\pi^2 \rho}{m^3}v_ff(v_f)\;\text{sinc}(mv_fd),
\end{equation}
where $d\equiv|\boldsymbol{d}|$ and $\text{sinc}(x) \equiv \sin(x)/x$. 
In cases with a general velocity distribution, analytic results are not available and the integral needs to be evaluated numerically.

Another useful quantity is $S_{\Phi}(f,\hat{k})$, the ``skymap" of ULDM field, which is the projection of $S_{\Phi}(f,\boldsymbol{d})$ onto the unit direction vector $\hat{k}$. We derive the explicit expression of $S_{\Phi}(f,\hat{k})$ in Appendix~\ref{appendix DM skymap}.

\section{Detect ULDM with Space-based Gravitational-Wave Interferometers} \label{Detect ULDM}

\subsection{Response of interferometer to ULDM}
Now, we discuss how ULDM is responded in a GW laser interferometer in space. We first define some useful quantities for the ULDM with mass $m$, including the Compton frequency, coherence time, and coherence length,
\begin{align}
    f_c &= \frac{mc^2}{2\pi \hbar} \approx 2.42 \times \left(\frac{m}{10^{-17}~\text{eV}}\right) ~\text{mHz}, \\
    \label{tc}\tau_c &= \frac{\hbar}{mc^2\sigma^2} \approx 6.58 \times 10^7 \left(\frac{10^{-17}~\text{eV}}{m}\right)~\text{s},\\
    \label{lc}\lambda_c &= \frac{\hbar}{m\sigma c} \approx 1.97 \times 10^{10} \left(\frac{10^{-17}~\text{eV}}{m}\right)~\text{km},
\end{align}
where $\sigma \sim 10^{-3}$ is the velocity dispersion of the local DM.
The sensitivity band of space-based GW detectors, such as LISA and Taiji, is around several mHz, which corresponds to the ULDM with a mass around $10^{-17}$~eV. Consequently, the detectable ULDM will have $\tau_c > T \sim 3 \times 10^7~\text{s}$~\footnote{While this criterion is satisfied for the majority of the sensitive band, the high frequency tail will deviate from this condition. To derive the precise sensitivity of the high frequency part, one can utilize the general framework presented in~\ref{Network}.} 
and $\lambda_c \gg L \sim 10^6~\text{km}$, where $T$ is the total observation time and $L$ is the arm length of interferometer. 
This also indicates that such detectors will not have sufficient frequency resolution to resolve the ULDM signal spectrum, which has a characteristic width of about $1/\tau_c$ in the Fourier domain. Thus, we can treat the ULDM as monochromatic, and Eq.~(\ref{dPhi disc}) can be approximated by
\begin{equation} \label{dPhi disc app}
     \nabla\Phi(x) \simeq -i\sqrt{\rho} e^{imt} \sum_{n}\sum_{l}  \sqrt{\Delta_{nl}}v_n\beta_{nl} 
    e^{-i\left( \boldsymbol{k}_{nl}\cdot\boldsymbol{x} - \theta_{nl}\right)}\;\hat{e}_l.
\end{equation}

To proceed, we need to specify the coupling between ULDM and the SM.
Here, we illustrate the physical effects with scalar ULDM, which is coupled with test mass in GW interferometer through the trace of energy-momentum tensor. Note that our methodology can be readily extended to vector and tensor ULDM with different couplings.
The action of test mass is given by~\cite{PhysRevD.100.123512}
\begin{equation}
    S = -\int m(\Phi) \sqrt{-\eta_{\mu \nu} dx^{\mu} dx^{\nu}},
\end{equation}
from which we can obtain the non-relativistic equation of motion, 
\begin{equation} \label{eom}
    \frac{d^2\boldsymbol{x}}{dt} \simeq -\kappa \alpha(\Phi) \nabla\Phi,\qquad
    \alpha(\Phi) \equiv \frac{\partial\ln\left[\kappa m(\Phi)\right]}{\partial(\kappa\Phi)},
\end{equation}
where $\kappa \equiv \sqrt{4\pi G}$ is the inverse of the Planck mass. The dimensionless factor $\alpha(\Phi)$ is determined by the interactions between the ULDM and the SM.
In this work, we consider the general parametrized interaction~\cite{ PhysRevD.82.084033},
\begin{equation}
    \mathcal{L}_{\Phi-\text{SM}}
    =\kappa \Phi \left[\frac{d_e}{4e^2}F_{\mu\nu}F^{\mu\nu} 
    - \frac{d_{g}\beta_{3}}{2g_{3}}G^{A}_{\mu\nu}G^{A \mu \nu}
    -\sum_{i=e,u,d}(d_{m_{i}}+\gamma_{m_{i}}d_{g})m_{i}\bar{\psi}_{i}\psi_{i}\right],
\end{equation}
where $d_{g}$ and $d_e$ denote for the couplings of the ULDM to the electromagnetic and gluonic field terms, $d_{m_{i}}$ ($i=e,u,d$) the couplings to the fermionic mass terms, 
$F_{\mu\nu}$, $G^A_{\mu \nu}$ are the electromagnetic and gluon field strength tensors,
$e$ the electron charge, $g_{3}$ the $\text{SU}(3)$ gauge coupling, 
$\beta_{3}$ the QCD beta function, and $\gamma_{m_{i}}$ the anomalous dimensions of fermion.
Around the energy scale of nuclei, $\alpha(\Phi)$ is approximately given by~\cite{ PhysRevD.82.084033}
\begin{equation} \label{alpha}
    \alpha(\Phi) \simeq d_g^* + \left[\left(d_{\hat{m}}-d_g\right) Q_{\hat{m}}^{\prime} + d_e Q_e^{\prime}\right],
\end{equation}
where 
\begin{equation}
    d^*_g \simeq d_g + 9.3\times10^{-2}(d_{\hat{m}}-d_g) + 2.7\times10^{-4}d_e,\qquad
    d_{\hat{m}} \equiv \frac{d_{m_d}m_d+d_{m_u}m_u}{m_d+m_u}, 
\end{equation}
The ``dilaton charges" $Q_{\hat{m}}^{\prime}$ and $Q_e^{\prime}$ carried by test mass are related to its composition through
\begin{equation}
    Q_{\hat{m}}^{\prime} = -\frac{0.036}{A^{1/3}} -1.4\times10^{-4}\frac{Z(Z-1)}{A^{4/3}},\qquad
    Q_e^{\prime} = 7.7\times10^{-4}\frac{Z(Z-1)}{A^{4/3}},
\end{equation}
where $Z$ denotes the atomic number and $A$ the mass number of the material. 
For the test masses of LISA and Taiji, which are made of gold-platinum alloy, we adopt $A=195.1$ and $Z=78$, respectively.

The velocity variation of test mass induced by the oscillation of the ULDM field can be obtained by integrating Eq.~(\ref{eom}),
\begin{equation}
    \frac{d\boldsymbol{x}}{dt} = -\frac{\lambda}{im} \nabla\Phi,
\end{equation}
where $\lambda \equiv \kappa\alpha(\Phi)$.
Then, the frequency modulation of laser in a single link due to the velocity variation~(Doppler shift) is 
\begin{equation} \label{1link}
    \eta_{rs}(t) \equiv \frac{\delta\nu}{\nu_0} = -\hat{n}_{rs} \cdot \left(\frac{d\boldsymbol{x}}{dt}\bigg|_{t,\boldsymbol{x}_r} - \frac{d\boldsymbol{x}}{dt}\bigg|_{t-L,\boldsymbol{x}_s}\right) 
    = \frac{\lambda}{im} \hat{n}_{rs} \cdot \left(\nabla\Phi\bigg|_{t,\boldsymbol{x}_r} - \nabla\Phi\bigg|_{t-L,\boldsymbol{x}_s}\right),
\end{equation}
where $\hat{n}_{rs}$ ($r,s = 1,2,3$) is the unit vector pointing from the sender spacecraft $s$ at position $\boldsymbol{x}_s$ to the receiver spacecraft $r$ at $\boldsymbol{x}_r$.
Substituting Eq.~(\ref{dPhi disc app}) into Eq.~(\ref{1link}), we have
\begin{equation} \label{1link sub}
    \begin{aligned}
    \eta_{rs}(t) &= -\lambda\frac{\sqrt{\rho}}{m} \hat{n}_{rs} \cdot 
    \left[ e^{imt} \sum_{n,l} \sqrt{\Delta_{nl}}v_n\beta_{nl} 
     e^{-i\left(\boldsymbol{k}_{nl}\cdot\boldsymbol{x}_r - \theta_{nl}\right)}\;\hat{e}_l \right.\\
     &\quad\quad\quad\quad\quad\quad\quad\quad \left.-e^{im(t-L)} \sum_{p,q} \sqrt{\Delta_{pq}}v_p\beta_{pq} 
     e^{-i\left(\boldsymbol{k}_{pq}\cdot\boldsymbol{x}_s - \theta_{pq}\right)}\;\hat{e}_q  \right] \\
     &\simeq -\lambda\frac{\sqrt{\rho}}{m} e^{imt}\left(1-e^{-imL}\right) 
     \sum_{n,l} \sqrt{\Delta_{nl}}v_n 
     \beta_{nl}e^{-i\left(\boldsymbol{k}_{nl}\cdot\boldsymbol{x}_r - \theta_{nl}\right)} \;\hat{e}_l\cdot\hat{n}_{rs} \;,
    \end{aligned}
\end{equation}
where we use the small antenna approximation $L \ll \lambda_c$ in the last line.
\begin{figure}
    \centering
    \includegraphics[width=0.50\linewidth]{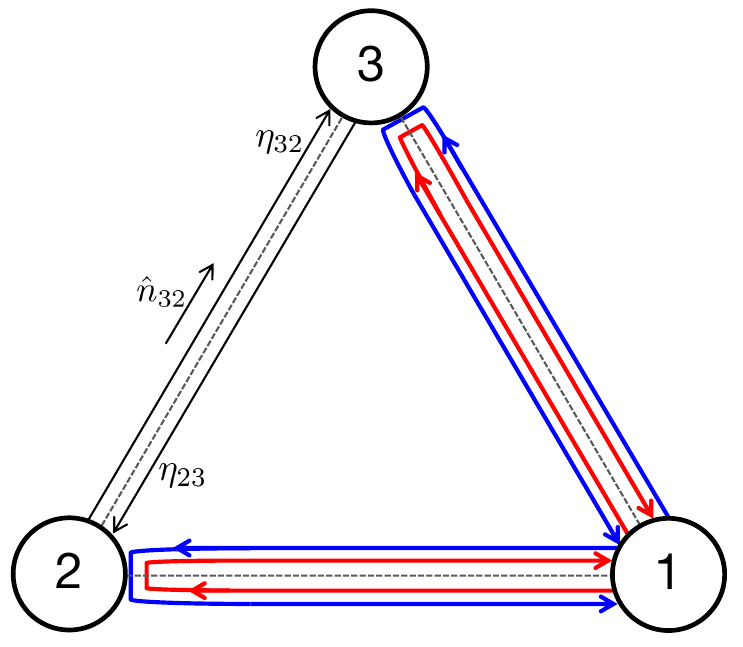}
    \caption{Illustration of the equilateral triangular configuration of space-based GW detectors such as LISA, Taiji and TianQin. The red and blue lines indicate the synthetic light path of TDI $X$ channel.}
    \label{fig:TDI X}
\end{figure}

From the single-link response Eq.~(\ref{1link sub}), we can then construct the round-trip signal $\xi_{rsr}(t)$, Michelson interference signal $M(t)$ and TDI $X$ channel signal $X(t)$. Their expressions are listed below:
\begin{align}
    \xi_{rsr}(t) &= \eta_{rs}(t)+\eta_{sr}(t-L) \notag\\ 
    &\simeq -\lambda\frac{\sqrt{\rho}}{m} e^{imt}\left(1-e^{-imL}\right)^2 
    \sum_{n,l} \sqrt{\Delta_{nl}}v_n\beta_{nl}
    e^{-i\left(\boldsymbol{k}_{nl}\cdot\boldsymbol{x}_r - \theta_{nl}\right)} \;\hat{e}_l\cdot\hat{n}_{rs}, \\
    
    \label{Michelson interf} M(t) &= \xi_{121}(t)-\xi_{131}(t) \notag\\
    &\simeq -\lambda\frac{\sqrt{\rho}}{m} e^{imt}\left(1-e^{-imL}\right)^2 
    \sum_{n,l} \sqrt{\Delta_{nl}}v_n\beta_{nl} e^{-i\left(\boldsymbol{k}_{nl}\cdot\boldsymbol{x}_1 - \theta_{nl}\right)}
    \left[\hat{e}_l\cdot(\hat{n}_{12}-\hat{n}_{13})\right], \\
    
    \label{tdiX} X(t) 
    &= \left[\xi_{131}(t)+\xi_{121}(t-2L)\right]-\left[\xi_{121}(t)+\xi_{131}(t-2L)\right] \notag\\
    &\simeq \left(1-e^{-2imL}\right)M(t)\;.
\end{align}

\subsection{The power spectral density of ULDM signal} \label{PSD of ULDM}
The ULDM signal is characterized statistically by the corresponding PSD.
For the Michelson interference, the auto-correlation function is given by
\begin{equation}
    R_M(\tau) \equiv \langle M(t)M(t+\tau)\rangle 
    = \frac{16\rho\lambda^2}{m^2}\sin^4\left(\frac{mL}{2}\right)I_{\text{vel}}\cos\left(m\tau\right) ,
\end{equation}
where the velocity integral $I_{\text{vel}}$ is defined as
\begin{equation} \label{velocity integral}
    I_{\text{vel}} \equiv \int d^3v\;f(\boldsymbol{v})\left[\boldsymbol{v}\cdot(\hat{n}_{12}-\hat{n}_{13})\right]^2,
\end{equation}
which measures the average kinetic energy of DM particles projected onto the sensitive axis $\hat{n}_{32} = \hat{n}_{12}-\hat{n}_{13}$ of detector.
From Eq.~(\ref{tdiX}), the auto-correlation function of $X(t)$ follows that
\begin{equation}
    R_X(\tau) \equiv \langle X(t)X(t+\tau)\rangle = 4\sin^2\left(mL\right)R_M(\tau).
\end{equation}
The two-sided PSDs are obtained by Fourier transforming the auto-correlation function. 
The results are
\begin{equation} \label{PSD M}
\begin{aligned} 
    S_M(f) = \frac{8\rho\lambda^2}{m^2}\sin^4\left(\frac{mL}{2}\right)I_{\text{vel}} \times
    \delta\left(f-f_c\right)
    \simeq \frac{2\rho\lambda^2}{\pi^2 f_c^2}\sin^4\left(\pi f_cL\right)I_{\text{vel}}T,
\end{aligned}
\end{equation}
where we replace the $\delta$ function by the finite observation time $T$, and similarly
\begin{equation} \label{PSD X}
    S_X(f) = 4\sin^2\left(2\pi fL\right)S_M(f).
\end{equation}
From Eq.~(\ref{PSD M}), it is clear that the ULDM signal will manifest as a monochromatic signal at the Compton frequency $f_c$ with its intensity being proportional to $T$.

For later convenience, we rewrite Eqs.~(\ref{PSD M}, \ref{PSD X}) in terms of the response function of TDI channel~\cite{PhysRevD.108.083007} and $S_{\Phi}(f,\boldsymbol{d})$.
Observe that for $T \ll \tau_c$, we have
\begin{equation}
    \frac{\nabla_{\boldsymbol{d}}}{-im} S_{\Phi}(f,\boldsymbol{d}) \simeq 
    \frac{\rho}{m^2} \int d^3v \; \boldsymbol{v} f(\boldsymbol{v}) e^{-im\boldsymbol{v}\cdot\boldsymbol{d}}
    \times\frac{1}{2}\delta\left(f-f_c\right).
\end{equation}
This suggests that we can replace $\boldsymbol{v}$ in Eq.~(\ref{velocity integral}) by the operator $\nabla_{\boldsymbol{d}}/(-im)$ and Eq.~(\ref{PSD M}) can be rewritten as 
\begin{equation}
    S_M(f) = \lambda^2 \big[\Gamma_M(f)S_{\Phi}(f,\boldsymbol{d})\big] \Big|_{\boldsymbol{d}=\boldsymbol{0}},
\end{equation}
where 
\begin{equation}
    \Gamma_M(f) =16\sin^4\left(\pi fL\right)
    \left[(\hat{n}_{12}-\hat{n}_{13}) \cdot \frac{\nabla_{\boldsymbol{d}}}{-im}\right]^2,
\end{equation}
which is just the square of modulus of the response function of the Michelson interference derived in Ref.~\cite{PhysRevD.108.083007}. In general, the two-sided PSD of the ULDM signal in a general TDI channel $O$ can be written as 
\begin{equation} \label{S_o = gamma S_phi}
    S_O(f) = \lambda^2 \big[\Gamma_O(f)S_{\Phi}(f,\boldsymbol{d})\big] \Big|_{\boldsymbol{d}=\boldsymbol{0}},
\end{equation}
where $\Gamma_O$ is the square of modulus of the response function of $O$ channel.

\subsection{Velocity integral} \label{Velocity integral}
The information about the phase space of DM is encoded into the velocity integral $I_{\text{vel}}$.
Here, we list the results of integral for three interested velocity distributions for later use.
\begin{itemize}
    \item The isotropic Maxwellian distribution:
    \begin{equation} \label{Maxwell}
        f(\boldsymbol{v}) = \frac{1}{(2\pi \sigma^2)^{3/2}} e^{-\frac{v^2}{2\sigma^2}},
    \end{equation}
     with the expectation value of velocity square $\langle v^2\rangle=3\sigma^2$ and $I^{\text{iso}}_{\text{vel}}=\sigma^2$.

     \item The $\delta$ distribution corresponds to the monochromatic plane waves:
     \begin{equation}
         f(\boldsymbol{v}) = \frac{1}{4\pi \sigma^2} \delta\left(v-\sigma\right),
     \end{equation}
     with $\langle v^2\rangle=\sigma^2$ and $I^{\delta}_{\text{vel}}=\sigma^2/3$.

     \item The boosted Maxwellian distribution in the Solar System Barycenter frame~(SSB):
     \begin{equation} \label{boosted Maxwell}
         f(\boldsymbol{v}) = 
         \frac{1}{(2\pi \sigma^2)^{3/2}} e^{-\frac{(\boldsymbol{v}+\boldsymbol{v}_{\odot})^2}{2\sigma^2}},
     \end{equation}
      where $\boldsymbol{v}$ is the velocity of DM particles in the SSB frame, and $\boldsymbol{v}_{\odot}$ is the velocity of the solar system in the galactic rest frame and $|\boldsymbol{v}_{\odot}|\simeq \sigma$.
      $I^{\odot}_{\text{vel}}=\sigma^2(1+\cos^2\iota)$, where $\iota$ is the angle between $\boldsymbol{v}_{\odot}$ and the sensitive axis $\hat{n}_{32}$.
\end{itemize}
Note that the relative magnitudes of the velocity integrals: $I^{\odot}_{\text{vel}} \ge I^{\text{iso}}_{\text{vel}} > I^{\delta}_{\text{vel}}$. Also, besides the isotropic background contribution $\sigma^2$, $I^{\odot}_{\text{vel}}$ get an additional term $\sigma^2\cos^2\iota$ due to the boost of reference frame. When $\iota=0$ or $\pi$, i.e.
$\hat{n}_{32}$ aligned or anti-aligned with the velocity of the Sun in the galactic rest frame, the signal power will get a factor of 2 enhancement compared to the isotropic case.

\section{Detection with A Single Detector} \label{Single detector}
In this section, we discuss the detection of ULDM with a single detector.
First, we examine the sensitivity of a single detector, considering the stochastic nature of the field. Then, we discuss how to distinguish the quasi-monochromatic ULDM signal from other monochromatic signals within the band.

\subsection{Sensitivity of A Single Detector} \label{Sensitivity of Single Detector}
Since the ULDM signal is contained entirely in the frequency bin with width $1/T$ centered at the Compton frequency of ULDM,
the likelihood of data $\tilde{d}=\tilde{s}+\tilde{n}$, where $\tilde{s}$ and $\tilde{n}$ stand for the Fourier amplitude of signal and noise, is given by \cite{PhysRevA.97.042506} 
\begin{equation} \label{1d stoch amp pdf}
    \mathcal{L}_{\text{stoch}}\left(\tilde{d}\Big|S_O(f_c), N_O(f_c)\right)
    =\frac{1}{\pi T\left(S_O+N_O\right)}\exp\left[-\frac{|\tilde{d}|^2}{T\left(S_O+N_O\right)}\right],
\end{equation}
where $S_O(f)$ and $N_O(f)$ are the two-sided PSDs of the ULDM signal and the detector noise in $O$ channel, respectively. $T$ is the observation time. Notice that the integral volume element of Eq.~$(\ref{1d stoch amp pdf})$ is $d\Re[\tilde{d}]d\Im[\tilde{d}]$, the real part and imaginary part of $\tilde{d}$.

We choose the excess power $P=|\tilde{d}|^2/\left(TN_O\right)$ as the detection statistic. The likelihood of $P$ can be obtained from Eq.~$(\ref{1d stoch amp pdf})$ through a variable transformation, 
\begin{equation} \label{1d stoch power pdf}
    \mathcal{L}_{\text{stoch}}\left(P\Big|S_O, N_O\right)
    =\frac{N_O}{S_O+N_O} \exp\left[-\frac{N_O}{S_O+N_O}P\right].
\end{equation}
The sensitivity of a detector is defined as the minimum value of $\lambda$
that can be detected with a detection probability $\gamma$ under a false alarm rate $\alpha$~\cite{Kay2001FundamentalsOS},
\begin{equation} \label{detection rate}
    \gamma = \int^{\infty}_{P_*(\alpha)} dP\;\mathcal{L}_{\text{stoch}}\left(P\Big|S_O(\lambda_{\min}), N_O\right),
\end{equation} 
where the threshold value $P_*(\alpha)$ is related to the false alarm rate $\alpha$ through
\begin{equation} \label{false alarm rate}
    \alpha = \int^{\infty}_{P_*} dP\;\mathcal{L}_{\text{stoch}}\left(P\Big|\lambda=0, N_O\right).
\end{equation}
If we observe $P>P_*$ in a specific realization of experiment, we conclude that the presence of ULDM signal; if we observe $P<P_*$, we report the null result. 

Substituting the likelihood Eq.~(\ref{1d stoch power pdf}) into Eqs.~(\ref{detection rate}, \ref{false alarm rate}), we have 
\begin{equation} \label{1d power ratio}
    S_O(\lambda_{\min}) = \left(\frac{\ln\alpha}{\ln\gamma}-1\right)N_O,
\end{equation}
or written in terms of the response function and the PSD of the field using Eq.~(\ref{S_o = gamma S_phi})
\begin{equation} \label{1d lambda^2}
    \lambda_{\min}^2  = \frac{N_O}{\Gamma_O S_{\Phi}}\left(\frac{\ln\alpha}{\ln\gamma}-1\right).
\end{equation}
As an illustration, we derive the sensitivity of $X$ channel to $d_g$ assuming that all other coupling parameters are zero.
Substituting Eqs.~(\ref{PSD M}, \ref{PSD X}) into Eq.~(\ref{1d power ratio}) and setting $d_{\hat{m}}=d_e=0$ in Eq.~(\ref{alpha}), we have
\begin{equation} \label{1d stoch sensitive}
    d^{\text{stoch}}_g \simeq 1.1\sqrt{\frac{m^2 N_X}{32\rho\kappa^2\sin^2(mL)\sin^4\left(mL/2\right)I^{\odot}_{\text{vel}}T}
    \left(\frac{\ln\alpha}{\ln\gamma}-1\right) },
\end{equation}
where $\ln\alpha/\ln\gamma \approx 58.4$ for $(\alpha,\gamma)=(0.05,0.95)$~\footnote{The choice of values for $\alpha$ and $\gamma$ is somewhat arbitrary and depends on the levels of type I and type II error one can tolerate~\cite{Jaranowski:2005hz}. Here, for a reliable detection, we prefer to choose a large $\gamma$ with a small $\alpha$, and adopt $(\alpha,\gamma)=(0.05,0.95)$ as our fiducial values.}, and the factor $1.1$ accounts for the difference between $d_g$ and $\alpha$.

For comparison, we rederive the sensitivity curve in the deterministic case presented in Ref.~\cite{PhysRevD.108.083007} with this frequentist formalism. The likelihood of the excess power $P$ in the deterministic case is given by~\cite{Centers:2019dyn}
\begin{equation} \label{1d det power pdf}
    \mathcal{L}_{\text{det}}\left(P\Big|S_O, N_O\right) = 
    e^{-\left(P+P_s\right)}I_0\left(2\sqrt{PP_s}\right),
\end{equation}
where $P_s\equiv|\tilde{s}|^2/TN_O$, $|\tilde{s}|^2/T = S_O[I^{\delta}_{\text{vel}}]$, and $I_0$ is the modified Bessel function of the first kind.
Following the same procedure outlined above, we have
\begin{equation} \label{1d det sensitive}
    d_g^{\text{det}} \simeq 1.1\sqrt{\frac{3m^2N_X}{32\rho\kappa^2\sigma^2\sin^2(mL)\sin^4\left(mL/2\right)\;T} F(\alpha,\gamma)},
\end{equation}
where $F(\alpha,\gamma) \approx 7.72$ for $(\alpha,\gamma)=(0.05,0.95)$. 
\begin{figure}
    \centering
    \includegraphics[width=0.72\linewidth]{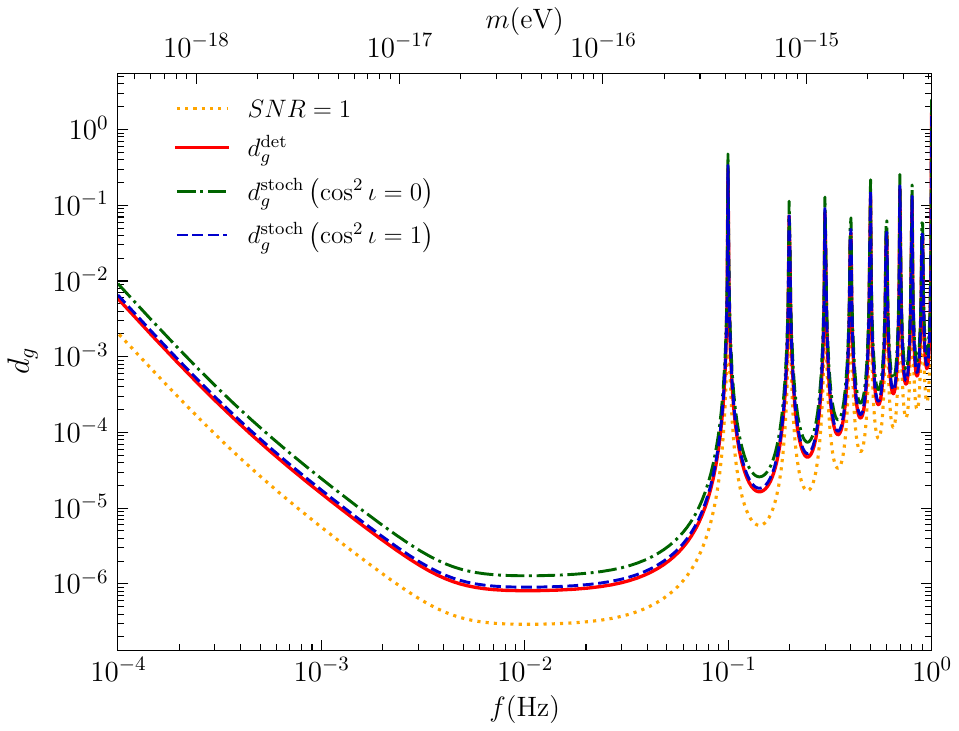}
    \caption{Comparison of sensitivities to $d_g$ in the deterministic and stochastic cases: the stochastic case with the boosted velocity distribution~(dash) or isotropic distribution~(dot dash), the deterministic case~(solid), and the $\text{SNR}=1$ estimation~(dot). Except for the $\text{SNR}=1$ case, the other three sensitivities correspond to a $95\%$ detection rate and a $5\%$ false alarm rate.}
    \label{fig:sensitivity compare}
\end{figure}

We present the sensitivities of a single detector in the deterministic and stochastic cases with a $95\%$ detection rate and a $5\%$ false alarm rate in Fig.~\ref{fig:sensitivity compare}. For illustration purposes, we adopt the noise parameters of Taiji, which are listed in Appendix~\ref{noise psd}.
As shown in Fig.~\ref{fig:sensitivity compare}, the minimum value of $d_g^{\text{det}}$ derived here is greater than that derived in Ref.~\cite{PhysRevD.108.083007} (denoted by $\text{SNR}=1$). This discrepancy arises because the previous sensitivity curve was a rough estimation obtained by setting the signal-to-noise ratio~(SNR) equal to one, and it only corresponds to a $22\%$ detection rate for $\alpha=0.05$ within this frequentist formalism.

Comparing Eqs.~(\ref{1d stoch sensitive}, \ref{1d det sensitive}), we observe that the ratio between the sensitivities in the deterministic and stochastic cases can be written as
\begin{equation} \label{C ratio}
    \left(\frac{d_g^{\text{stoch}}}{d_g^{\text{det}}}\right)^2 = \frac{I^{\delta}_{\text{vel}}}{I^{\odot}_{\text{vel}}} \times C(\alpha,\gamma),
\end{equation}
where $C(\alpha,\gamma) = \left(\ln\alpha/\ln\gamma-1\right)/F(\alpha,\gamma)$
characterizes the reduction in sensitivity due to the statistical difference in signal model between the two cases,
and the ratio $I^{\delta}_{\text{vel}}/I^{\odot}_{\text{vel}}$ compares the average kinetic energy of DM particles projected onto the sensitive axes in the two cases.
Whether the sensitivity in the stochastic case is degraded compared to the deterministic case is determined by the relative magnitude of the two terms on the right-hand side of Eq.~(\ref{C ratio}). 
For example, when $(\alpha, \gamma) = (0.05, 0.95)$ and $\cos^2\iota=1$,  we have $C\approx7.4 > I^{\odot}_{\text{vel}}/I^{\delta}_{\text{vel}}=6$ and $d_g^{\text{stoch}}\approx1.11d_g^{\text{det}}$. However, for larger value of $\alpha$ and
smaller value of $\gamma$, the statistical difference in signal model becomes less significant, and the value of $C$ decreases. When $C(\alpha, \gamma) < 6$, satisfied by $(\alpha, \gamma) = (0.10, 0.90)$, the reduction due to the statistical effect is entirely compensated, resulting in even better sensitivity in the stochastic case.

We have noticed that there is discrepancy between our result and the literature.
It has been pointed out that the stochastic effect will typically relax constraints or sensitivities by a factor ranging from $3$ to $10$, depending on experimental details, compared to the deterministic case if the experiments probe the field $\Phi(x)$ directly~\cite{Centers:2019dyn}. Additionally, a factor of $2$ was found for the gradient coupling under certain assumptions in the supplementary material of Ref.~\cite{Centers:2019dyn}. 
The discrepancy mainly arises from two factors.
First, we adopt a different random phase model. As shown in Appendix~\ref{Random phase model}, the two random phase models predict different strengths of the gradient correlation even for the same velocity distribution, resulting in an $\mathcal{O}(1)$ difference in the signal strength in the stochastic case. 
Secondly and more importantly, when evaluating the sensitivity in the deterministic case, we effectively average the signal over the sky to account for the randomness in the direction of the field gradient, as well as the evolution of the sensitive axis due to the spacecrafts' orbital motion. This manifests as the $1/3$ factor showing in $I^{\delta}_{\text{vel}}=I^{\text{iso}}_{\text{vel}}/3$.
In contrast, in Ref.~\cite{Centers:2019dyn}, the sensitive axis is assumed to align with the mean direction of the DM wind, and a larger $I^{\delta}_{\text{vel}}$ was used. Thus, our signal in the deterministic case is weaker compared to that in Ref.~\cite{Centers:2019dyn}.
More details can be found in Appendix~\ref{Random phase model}.

\begin{figure}
    \centering
    \includegraphics[width=0.75\linewidth]{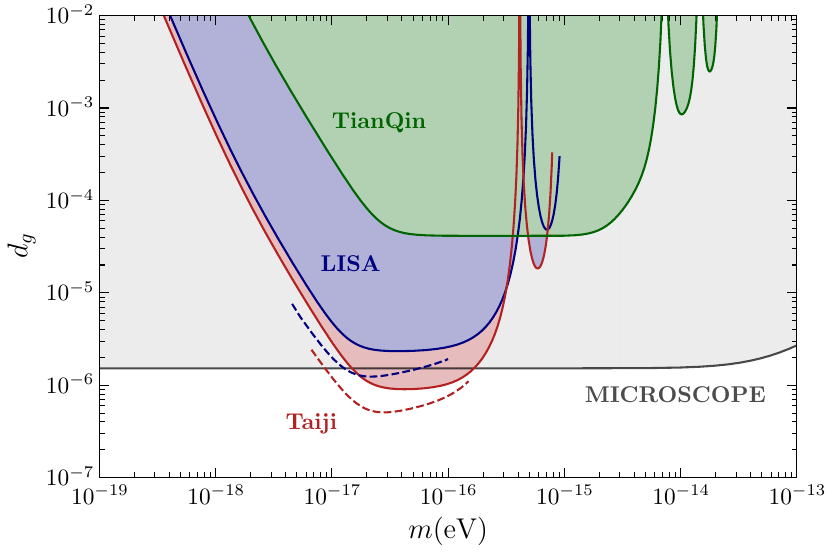}
    \caption{The projected sensitivities to $d_g$ of space-based GW detectors with a $95\%$ detection rate and a $5\%$ false alarm rate. The solid lines represent the one-year sensitivity of $X$ channel, while the dash lines are the four-year sensitivity of the optimal channels $A$, $E$, and $T$. See the discussion in~\ref{Sensitivity of Single Detector} for more details. Also shown is the $2\sigma$ constraint from the final results of MICROSCOPE~\cite{PhysRevLett.129.121102}. 
    Note that in this plot we set all the coupling parameters except $d_g$ to zero.}
    \label{fig:sensitivity}
\end{figure}
In Fig.~\ref{fig:sensitivity}, we present the projected sensitivities of space-based GW detectors LISA, Taiji and TianQin to $d_g$ with a $95\%$ detection rate and a $5\%$ false alarm rate, along with the $2\sigma$ constraint from the final results of MICROSCOPE~\cite{PhysRevLett.129.121102}. 
The noise parameters adopted for the three GW detectors are summarized in Appendix~\ref{noise psd}.
The solid lines correspond to the sensitivities of $X$ channel when the sensitive axis aligns with the direction of motion of the solar system in the galactic-rest frame, assuming a one-year observation time. As illustrated, with a one-year observation time, LISA will approach the constraint given by MICROSCOPE, and Taiji will improve the constraint by a factor of $1.7$ around $4.2\times10^{-17}$~eV, demonstrating the potential of GW detector to probe ULDM.

Considering the optimistic outlook for longer mission lifetimes, these one-year sensitivities are relatively conservative. We also plot the four-year sensitivities of the optimal TDI channels $A$, $E$, and $T$~\cite{PhysRevD.66.122002}. Compared to the one-year sensitivities, the four-year optimal sensitivities benefit from longer observation times~\footnote{The sensitivity is inversely proportional to the square root of the observation time if it is shorter than the coherence time of ULDM, as evident in Eq.~(\ref{1d stoch sensitive}), and it improves as $T^{-1/4}$ if $T\gg\tau_c$~\cite{PhysRevA.97.042506,PhysRevD.97.123006,nakatsuka2022stochastic}.}
and gain an additional $\sqrt{2}$ enhancement from the optimal channels~\cite{PhysRevD.108.083007}.
As shown in Fig.~\ref{fig:sensitivity}, with a four-year observation time, both LISA and Taiji will explore some unconstrained regions in parameter pace. More specifically, LISA will improve the constraint by a factor of $1.2$ around $2.2\times10^{-17}$~eV, while the improvement from Taiji is about $3.0$ around $2.8\times10^{-17}$~eV.

\begin{figure}[p]
    \centering
    \includegraphics[width=0.51\linewidth]{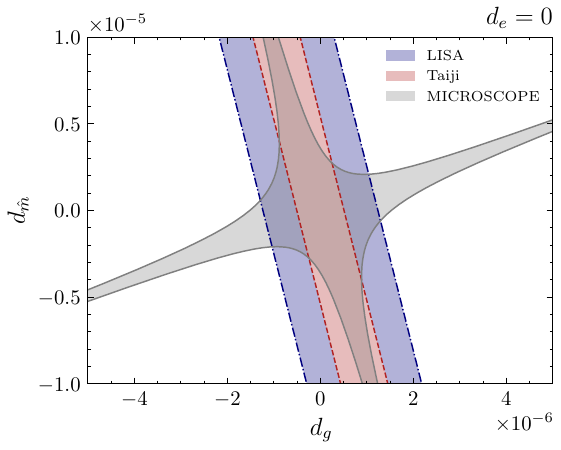}
    \includegraphics[width=0.51\linewidth]{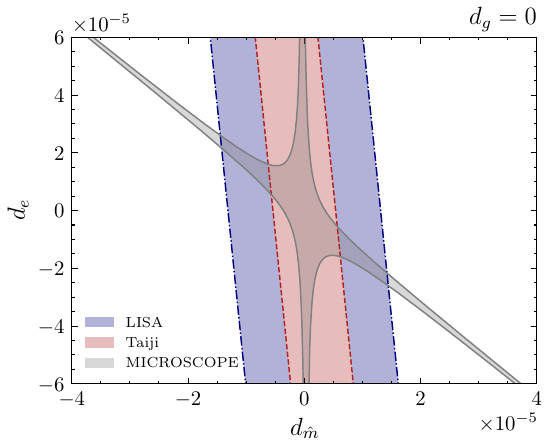}
    \includegraphics[width=0.51\linewidth]{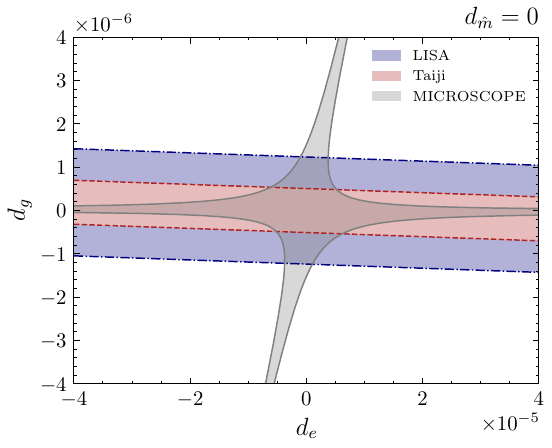}
    \caption{The 2D parameter regions to be probed by the four-year optimal sensitivity of LISA and Taiji at $m=2.2\times10^{-17}$~eV and $m=2.8\times10^{-17}$~eV, respectively, along with the $2\sigma$ constraint from MICROSCOPE~\cite{PhysRevLett.129.121102}.}
    \label{fig:dgdmde}
\end{figure}

Before ending this subsection, we emphasize that there is an essential difference on probing ULDM between GW experiments and the MICROSCOPE-like EP tests.
While MICROSCOPE probes a linear combination of coupling constants $\left([Q'_{\hat{m}}]_{\text{Pt}}-[Q'_{\hat{m}}]_{\text{Ti}}\right)d^*_g(d_{\hat{m}}-d_g) + \left([Q'_{e}]_{\text{Pt}}-[Q'_{e}]_{\text{Ti}}\right)d^*_gd_e$, GW experiments are sensitive to a different combination of coupling constants, see Eq.~(\ref{alpha}). Consequently, the two types of experiments probe different subspaces in parameter space. In Fig.~\ref{fig:dgdmde}, we compare the 2D contours probed by MICROSCOPE and GW experiments. These sensitive regions do not coincide, and 
there are scenarios, such as $d_g=d_{\hat{m}}$ and $d_e=0$, which evade constraints from MICROSCOPE but are limited by GW experiments. Therefore, EP tests and GW experiments complement each other.

\subsection{Discriminate ULDM Signal From Monochromatic GWs}
As elaborated in the preceding sections, ULDM will manifest as a characteristic sinusoidal oscillation in space-based GW interferometers. However, within the detector's frequency band, there are also other monochromatic signals, such as gravitational waves from galactic white dwarf binaries~(WDBs)~\cite{Sathyaprakash:2009xs}. 
This raises the question: How can we distinguish between a monochromatic signal originating from ULDM and one from WDBs?

One potential approach involves leveraging the distinct responses of TDI channels to ULDM and GWs. As shown in Ref.~\cite{PhysRevD.108.083007}, while $X$ channel exhibits similar responses to ULDM and GWs throughout the frequency band,  
the Sagnac combination $\zeta$ shows a weaker response to ULDM compared to GWs. 
Therefore, the ratio between the SNR of ULDM signal measured in $\zeta$ channel and that measured in $X$ channel will be smaller than the corresponding ratio of GWs.

Another distinguishing factor between ULDM signal and the GWs from WDBs lies in their spatial characteristics. Since space-based GW detectors primarily search for WDBs in the Milky Way, the GWs emitted by these sources predominantly emanate from the direction of the galactic disk. On the contrary, the direction of the ULDM field gradient is random. In an isotropic halo, the probability density function of the gradient's direction vector is uniform across the celestial sphere and there is a considerable probability that the direction vector does not align with the galactic disk. Therefore, if a monochromatic signal is detected and subsequent parameter estimation reveals that its sky location does not lie in the galactic disk, then it becomes more plausible that the signal originates from ULDM rather than WDB.

\section{Detection with A Detector Network} \label{Network}
Given the overlap in operation durations of several projected space-based GW missions~\cite{amaroseoane2017laser,Hu:2017mde,Luo_2016}, joint observations are possible. In this section, we discuss the detection of ULDM using a detector network, with a particular focus on identifying the optimal detector configuration. Although our framework can be readily extended to a multi-detector network, we restrict our discussion to a two-detector network here for the sake of simplicity.

\subsection{The general framework} \label{The general framework}
For the specific signals $\tilde{s}_1$ and $\tilde{s}_2$, the likelihood of data of two detectors with uncorrelated, stationary and Gaussian noise is given by
\begin{equation}
\begin{aligned}
    \mathcal{L}\left(\tilde{d}_1,\tilde{d}_2 \Big|\tilde{s}_1, \tilde{s}_2, N_1(f), N_2(f)\right) &= 
    \prod^{n}_{i=1} \frac{1}{\pi TN_1(f_i)} 
    \exp\left[-\frac{\left|\tilde{d}_1(f_i)-\tilde{s}_1(f_i)\right|^2}{TN_1(f_i)}\right] \\
    &\qquad\qquad\quad\times
    \prod^{n}_{j=1} \frac{1}{\pi TN_2(f_j)}
    \exp\left[-\frac{\left|\tilde{d}_2(f_j)-\tilde{s}_2(f_j)\right|^2}{TN_2(f_j)}\right]
\end{aligned}
\end{equation}
where $N_1(f)$ and $N_2(f)$ are the two-sided noise PSD of detectors, respectively, and the subscripts $i$ and $j$ label the frequency bins. In our scenario, $N_1(f)$ and $N_2(f)$ correspond to the noise PSD of TDI channels from two space-based GW detectors, like LISA and Taiji.
We can write the above likelihood in a more compact form,
\begin{equation}
    \mathcal{L}\left(\mathbf{\tilde{d}} \Big|\mathbf{\tilde{s}}, N_1, N_2\right) = 
    \frac{1}{\det\left[\pi T\mathbf{C}\right]}
    e^{-(\mathbf{\tilde{d}-\tilde{s}})^{\dagger}(T\mathbf{C})^{-1}(\mathbf{\tilde{d}-\tilde{s}})},
\end{equation}
where $\mathbf{\tilde{d}}=[\tilde{d}_1(f_1),\tilde{d}_2(f_1),\dots,\tilde{d}_1(f_n),\tilde{d}_2(f_n)]$, $\mathbf{\tilde{s}}=\left[\tilde{s}_1(f_1),\tilde{s}_2(f_1),\dots,\tilde{s}_1(f_n),\tilde{s}_2(f_n)\right]$, 
and 
$\mathbf{C}=\text{diag}\left[N_1(f_1),N_2(f_1),\dots,N_1(f_n),N_2(f_n)\right]$.

Given that the ULDM signal $\tilde{s}_1$ and $\tilde{s}_2$ are random variables and our interest lies in their statistical characteristics rather than a particular realization, we marginalize over $\mathbf{\tilde{s}}$
\begin{equation}
\begin{aligned}
    \mathcal{L}_{\text{stoch}}\left(\mathbf{\tilde{d}} \Big|\lambda, S_{\Phi}, N_1, N_2\right) = 
    \int d\mathbf{\tilde{s}} \;\mathcal{L}\left(\mathbf{\tilde{d}} \Big|\mathbf{\tilde{s}}, N_1, N_2\right)  \mathcal{L}\left(\mathbf{\tilde{s}}\Big|S_{\Phi}\right).
\end{aligned}   
\end{equation}
The marginalized likelihood is given by \cite{PhysRevD.87.122003} 
\begin{equation} \label{network likelihood}
    \mathcal{L}_{\text{stoch}}\left(\mathbf{\tilde{d}} \Big|\lambda, S_{\Phi}, N_1, N_2\right) = 
    \frac{1}{\det[\pi T\mathbf{\Sigma}]}
    e^{-\mathbf{\tilde{d}}^{\dagger}\left(T\boldsymbol{\Sigma}\right)^{-1}\mathbf{\tilde{d}}},
\end{equation}
where $\mathbf{\Sigma}=\text{diag}\left[\Sigma_1,\dots,\Sigma_n\right]$ is a block diagonal matrix with
\begin{equation} \label{signal+noise corl matrix}
    \Sigma_i = \begin{bmatrix}
        N_1(f_i) & \\
        & N_2(f_i)
    \end{bmatrix} 
    + \lambda^2\int d\hat{k}\begin{bmatrix}
        \chi_{11}(f_i,\hat{k},v)S_{\Phi}(f_i,\hat{k}) 
        &\;\chi_{12}(f_i,\hat{k},v)S_{\Phi}(f_i,\hat{k}) \\
        \chi_{21}(f_i,\hat{k},v)S_{\Phi}(f_i,\hat{k}) 
        &\;\chi_{22}(f_i,\hat{k},v)S_{\Phi}(f_i,\hat{k})
    \end{bmatrix},
\end{equation}
and 
\begin{equation} \label{chi_ab}
    \chi_{ab}(f,\hat{k},v) := F_a(f,\hat{k},v)F^*_b(f,\hat{k},v) e^{-i2\pi fv\hat{k}\cdot \Delta \mathbf{x}_{ab}},
\end{equation}
where $v$ is the velocity of ULDM particle, $F_a(f,\hat{k},v)$ is the response function of TDI channel $O$ of the $a\textit{th}$ detector, and $\Delta \mathbf{x}_{ab}=\mathbf{x}_a-\mathbf{x}_b$ with $\mathbf{x}_a$ and $\mathbf{x}_b$ the position vectors of two detectors, respectively. 
Given that $v$ is related to $f$ through the dispersion relation of massive particle $v=\sqrt{1-f^2_c/f^2}$, $\chi_{ab}$ can also been viewed as dependent on the mass rather than the velocity of ULDM particle. 

Analogous to SGWB detection~\cite{PhysRevD.59.102001,Romano:2016dpx}, we define the overlap reduction function~(ORF) of ULDM as
\begin{equation} \label{orf}
    \gamma_{ab}(f,v) := \int d\hat{k}\;\chi_{ab}(f,\hat{k},v),
\end{equation}
which quantifies the degree of coherence of the ULDM signals observed by two detectors. In contrast to the ORF defined in SGWB detection, the ORF of ULDM exhibits velocity dependency~(or equivalently mass dependent), arising from both the velocity-dependent response function and the exponential term. Consequently, it becomes necessary to construct a template bank of ORFs for different ULDM masses in the practical searches.
\begin{figure}
    \centering
    \includegraphics[width=0.6\linewidth]{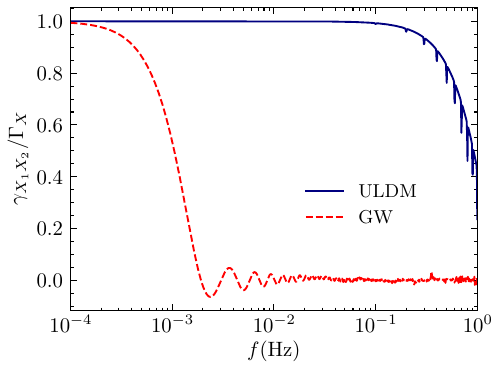}
    \caption{ The ratio $\gamma_{ab}(f,v)/\Gamma_{O}(f,v)$ measures the decrease in correlation.
    Here, we present the ratio for the two $X$ channels of a pair of identical detectors with an arm length $L=3\times 10^6$ km. 
    The two detectors are parallel to each other and separated by a distance of $|\Delta\mathbf{x}_{ab}|=348$ light-seconds, corresponding to the configuration where one detector leads the Earth by an angle of $20^{\circ}$, while the other trails the Earth by $20^{\circ}$. We set $v=\sigma\approx10^{-3}$.}
    \label{fig:orf_shift}
\end{figure} 

As evident from Eq.~(\ref{signal+noise corl matrix}), the marginalization introduces both the auto-correlation of $\tilde{d}_1$ and $\tilde{d}_2$ as well as the cross-correlation between $\tilde{d}_1$ and $\tilde{d}_2$, manifesting as additional diagonal and off-diagonal terms in $\Sigma_i$, respectively. The strength of cross-correlation is characterized by the ORF and the ``sky map" $S_{\Phi}(f,\hat{k})$, akin to SGWB detection.
However, unlike the SGWB scenario where the exponential term in Eq.~(\ref{chi_ab}) oscillates violently and $\gamma_{ab} (a\neq b)$ decays rapidly when $f>1/|\Delta \mathbf{x}_{ab}|$, the ORF of ULDM will decay when $|\Delta \mathbf{x}_{ab}|>2\pi\lambda_c \approx 10^3/f_c$ due to the longer wavelength of non-relativistic ULDM particles. Therefore, even if two detectors are separated by more than $1/f_c$ apart, they will still retain correlation for ULDM detection. 
Assuming a configuration where two identical detectors are co-aligned but separated by a distance of $|\Delta\mathbf{x}_{ab}|=348$ light seconds, which mimics the leading and trailing $20^{\circ}$ configuration of LISA and Taiji, we plot the ORF of ULDM and GW between the two $X$ channels in Fig.~\ref{fig:orf_shift}. We normalize the $\gamma_{X_1X_2}$ by $\Gamma_X$, and the ratio $\gamma_{X_1X_2}/\Gamma_X$ measures the decrease in correlation due to the separation between detectors. As illustrated, the GW ORF decays at much smaller frequencies than the ULDM ORF.

\begin{figure}[h]
    \centering
    \includegraphics[width=1\linewidth]{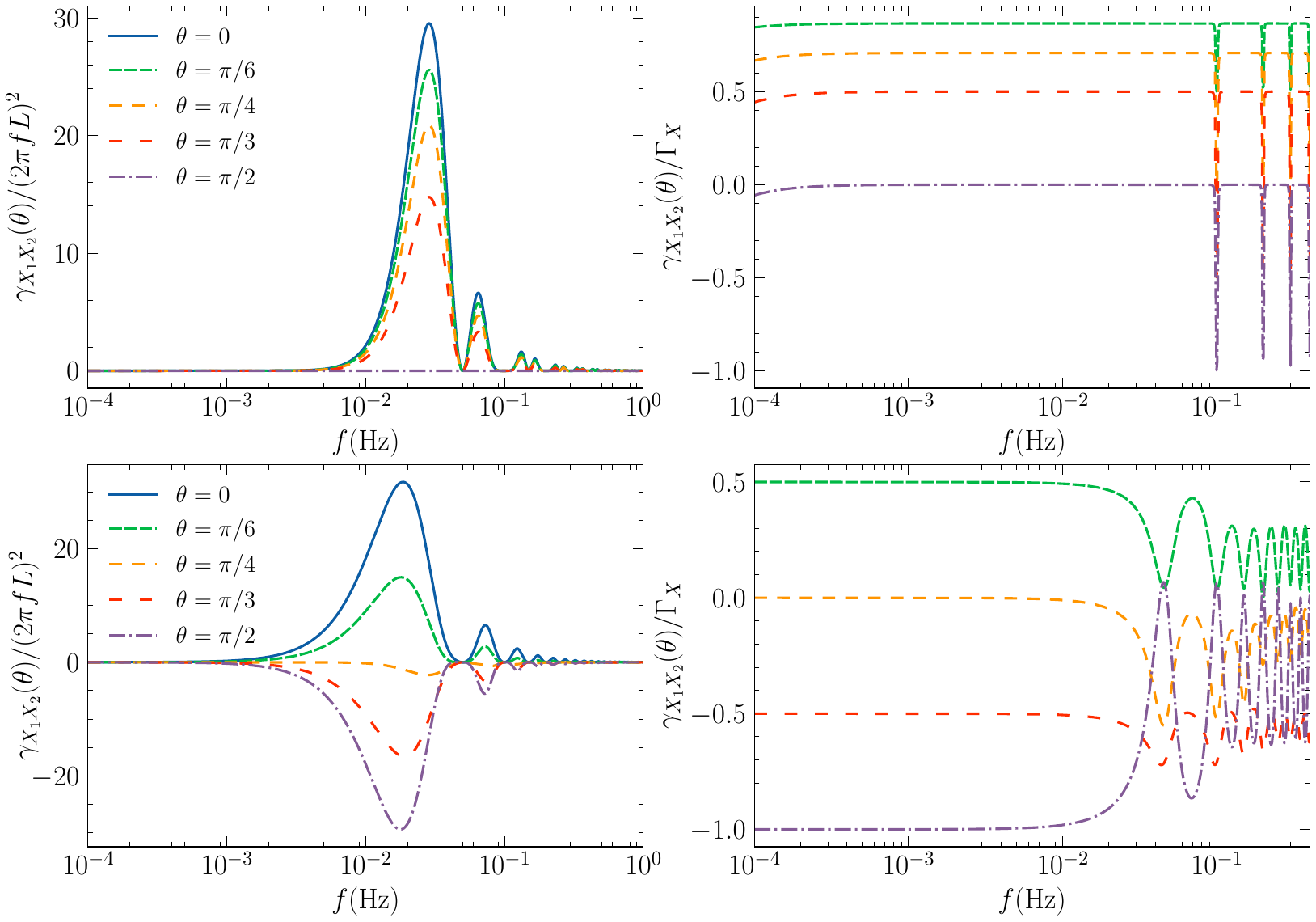}
    \caption{The overlap reduction function between the two $X$ channels of co-located identical detectors, where $\theta$ indicates the angle between their sensitive axes. The co-located and co-aligned configuration corresponds to $\theta=0$, and $\gamma_{X_1X_2}(\theta=0)=\Gamma_{X}$. The first row shows the ORF of ULDM. The case where the sensitive axes are orthogonal is represented by the dash-dot purple line. In this case, the ORF almost vanishes across the entire band, and the ULDM signals in the two detectors are uncorrelated. The second row shows the ORF of GW.}
    \label{fig:orf_rotation}
\end{figure}
In addition, the ULDM ORF has another interesting aspect, stemming from the directional dependence of the ULDM response function $F_a(f,\hat{k},v)$.
Due to the gradient coupling and the fact $L\ll\lambda_c$, $F_a(f,\hat{k},v)$ is highly direction-dependent. For non-co-aligned detectors, their directional responses mismatch, and the number of directions, for which the product $F_a(f,\hat{k},v)F^*_b(f,\hat{k},v)$ is vanishingly small, increases with the angle between the detectors' sensitive axes. Consequently, the ULDM ORF decreases monotonically as a function of the angle. In Fig.~\ref{fig:orf_rotation}, we plot the ORF between the two $X$ channels of co-located detectors, where $\theta$ denotes the angle between their sensitive axes~\footnote{Here, we restrict $\theta\in[0,\pi/2)$ since the ULDM ORF for $\theta\in[\pi/2, \pi)$ only differs by an overall minus sign.}. When $\theta=\pi/2$~(the dash-dot purple line), i.e. the sensitive axes are orthogonal, there is a minimal overlap between the detectors' responses and the ULDM ORF almost vanishes across the whole band~\footnote{Given the similarity between vector field and the gradient coupling of scalar field, we expect that there are similar results for the ORF of vector field.} 
Thus, in this case, the signals observed in the two $X$ channels are uncorrelated.
For the GW ORF, there is a similar behavior in the low frequency part when $\theta=\pi/4$. However, in the high frequency part, the GW response function depends on the direction in a more complex manner, and it is difficult for $F_a(f,\hat{k},v)F^*_b(f,\hat{k},v)$ with different $\hat{k}$ to cancel each other. Consequently, the GW ORF oscillates and deviates from zero in the high frequency part.

With the above formalism, we can derive the sensitivity of the network in the most general situation.
However, in our scenario, things simplify considerably. Since the ULDM signal is concentrated within the single frequency bin centered around its Compton frequency, we can safely ignore all other frequency bins containing only noise in the likelihood. Additionally, We assume identical detectors and use a isotropic spectrum $S_{\Phi}(f)$ to streamline our analysis such that analytic estimations are possible.
To simplify notation, we omit the subscript of $f_c$ and use $f$ to denote the frequency bin containing the Compton frequency in the following discussion.

\subsection{The co-aligned and co-located configuration} \label{The co-aligned and co-located config}
We first derive the sensitivity of the network in the co-aligned and co-located configuration,
where the signals in two detectors are strongly correlated. 
In this case, the exponential term in Eq.~(\ref{chi_ab}) equals one and the ORFs reduce to the square of modulus of the response function defined in \ref{PSD of ULDM}. For the two $X$ channels, we have
\begin{equation} \label{signal+noise corl matrix of optimal case}
    \Sigma = N(f)\begin{bmatrix}
        1 & \\
         & 1
    \end{bmatrix}
    +\lambda^2\Gamma_X(f) S_{\Phi}(f)\begin{bmatrix}
        1 & 1\\
        1 & 1
    \end{bmatrix},
\end{equation}
where $N(f)\equiv N_1(f)=N_2(f)$. 

We use the maximum likelihood estimator~(MLE) of $\lambda^2$ as the detection statistic~\footnote{
While maximum likelihood ratio~(MLR) usually is the optimal detection statistic for the hypothesis testing problem, here we use MLE to facilitate an analytical approach. We anticipate that MLR will give similar results, given the high SNR of signal. It would be prudent to revisit the question numerically using MLR through the Monte Carlo simulations. 
} and denote it by $\hat{\lambda}^2$. 
For a specific data $\mathbf{\tilde{d}}$, $\hat{\lambda}^2$ is determined by
\begin{equation} \label{ML estimator eq}
    \frac{\partial \ln\mathcal{L}_{\text{stoch}}}{\partial\lambda^2}\Big|_{\hat{\lambda}^2} = 0.
\end{equation}
Substituting Eqs.~(\ref{network likelihood}, \ref{signal+noise corl matrix of optimal case}) into Eq.~$(\ref{ML estimator eq})$, we have
\begin{equation}
   T\left(4\hat{\lambda}^2\Gamma_X S_{\Phi} + 2N\right) - \left(\tilde{d}_1+\tilde{d}_2\right)^*\left(\tilde{d}_1+\tilde{d}_2\right) = 0,
\end{equation}
or written in terms of the real and imaginary parts of $\tilde{d}_1$ and $\tilde{d}_2$,
\begin{equation} \label{optimal detec statis}
    \hat{\lambda}^2 = 
    \frac{\left(\Re[\tilde{d}_1]+\Re[\tilde{d}_2]\right)^2
    +\left(\Im[\tilde{d}_1]+\Im[\tilde{d}_2]\right)^2 - 2TN}{4T\Gamma_X S_{\Phi}}.
\end{equation}
Now we need to derive the probability density function~(PDF) of $\hat{\lambda}^2$ given the true value of coupling strength $\lambda$. In general, this can only be accomplished numerically through the Monte Carlo simulation. However, due to the specific form of the detection statistic Eq.~(\ref{optimal detec statis}) and the likelihood Eq.~(\ref{network likelihood}), we can obtain the PDF of $\hat{\lambda}^2$ analytically.

Since $\Sigma$ is a real symmetric matrix, $\Sigma^{-1}$ can be diagonalized through an orthogonal transformation. After diagonalization, the likelihood takes the form,
\begin{equation} \label{optimal network likelihood diag}
    \mathcal{L}_{\text{stoch}}\left(u_1, u_2, v_1, v_2\Big|\lambda, S_{\Phi}, N_1, N_2\right) = 
    \frac{1}{4\pi^2}\exp\left(-\frac{u^2_1+u^2_2+v^2_1+v^2_2}{2}\right),
\end{equation}
where the new variables are related to the old ones by
\begin{equation} 
    \begin{aligned}
        u_1 &= \frac{\Re[\tilde{d}_1]-\Re[\tilde{d}_2]}{\sqrt{TN}},\qquad 
        v_1 = \frac{\Re[\tilde{d}_1]+\Re[\tilde{d}_2]}{\sqrt{T\zeta}}, \\
        u_2 &= \frac{\Im[\tilde{d}_1]-\Im[\tilde{d}_2]}{\sqrt{TN}},\qquad 
        v_2 = \frac{\Im[\tilde{d}_1]+\Im[\tilde{d}_2]}{\sqrt{T\zeta}}, \\
    \end{aligned}
\end{equation}
where $\zeta\equiv2\lambda^2\Gamma_XS_{\Phi}+N$.
In this new set of variables, we have
\begin{equation}
    \hat{\lambda}^2 = \frac{\zeta(v_1^2+v_2^2) - 2N}{4\Gamma_X S_{\Phi}}.
\end{equation}
The term $v_1^2+v_2^2$ is the sum of the squares of two independent Gaussian variables with zero mean and unit variance, and it is well known that this sum follows the $\chi^2$ distribution with $2$ degrees of freedom.
Thus, the PDF of $\hat{\lambda}^2$ can be obtained from the $\chi^2$ distribution through a variable transformation. The result is
\begin{equation} \label{2d optimal lambda dis}
    \mathcal{L}_{\text{stoch}}\left(\hat{\lambda}^2\Big|\lambda, S_{\Phi}, N_1, N_2\right)
    = \frac{2\Gamma_X S_{\Phi}}{\zeta}\exp\left[-\left(\frac{2\Gamma_X S_{\Phi}}{\zeta}\hat{\lambda}^2+\frac{N}{\zeta}\right)\right].
\end{equation}

With the PDF of detection statistic in hand, we are now prepared to calculate the sensitivity of the network in the co-aligned and co-located configuration. The minimum value of $\lambda$ that can be detected, with a detection rate $\gamma$ and a false alarm rate $\alpha$, is determined by
\begin{align} 
    \label{alpha+gamma1} \alpha &= \int^{\infty}_{\hat{\lambda}^2_*} d\hat{\lambda}^2\; 
    \mathcal{L}_{\text{stoch}}\left(\hat{\lambda}^2\Big|\lambda=0, S_{\Phi}, N_1, N_2\right), \\
    \label{alpha+gamma2} \gamma &= \int^{\infty}_{\hat{\lambda}^2_*} d\hat{\lambda}^2\;
    \mathcal{L}_{\text{stoch}}\left(\hat{\lambda}^2\Big|\lambda_{\text{min}}, S_{\Phi}, N_1, N_2\right).
\end{align}
Substituting Eq.~$(\ref{2d optimal lambda dis})$ into Eqs.~(\ref{alpha+gamma1}, \ref{alpha+gamma2}), we have
\begin{equation} \label{lambda min strong corl}
    \lambda_{\min}^2 = \frac{N}{2\Gamma_X S_{\Phi}} \left(\frac{\ln\alpha}{\ln\gamma}-1\right).
\end{equation}
Comparing with Eq.~$(\ref{1d lambda^2})$, we observe a $\sqrt{2}$ improvement in sensitivity. In general, if there are $n$ co-aligned and co-located identical detectors, we will have a $\sqrt{n}$ improvement.

This is in sharp contrast with SGWB detection. In SGWB detection, we typically achieve about several orders of magnitude improvement in sensitivity when correlating the outputs of two detectors. It may seem surprising we only have a $\sqrt{2}$ improvement for ULDM. 
To address this question, we need to understand why correlation usually can lead to a substantial enhancement in sensitivity in SGWB detection. 

A key technique employed in the search of deterministic GW signals is matched filtering. In the context of ``deterministic", the signal is uniquely determined by its parameters, which allows us to predict its shape either in the time domain or frequency domain. Leveraging this predicted shape enables us to match it with the detector's output, thereby greatly improving sensitivity. However, for ``stochastic" signals, their shape in a specific realization cannot be predicted even with the knowledge of the parameters determining their statistical properties. To tackle this issue, we require the use of two detectors. With two detectors, we can use the output of one detector as the template to match the output of the other detector. This is precisely the correlation technique used in SGWB detection. 
The picture in the detection of ULDM is different. Although the amplitude of ULDM signal is stochastic, we know it consists of oscillating sine waves with frequency $f_c$. Using sine waves with different frequencies as templates, essentially performing the Fourier transform, will filter out the ULDM signal from the data. Therefore, we have performed matched filtering using the excess power approach in sec.~\ref{Sensitivity of Single Detector}. The $\sqrt{2}$ improvement in sensitivity when correlating the output of two detectors in ULDM detection corresponds to the $\sqrt{2}$ improvement when optimally combining the measurements of two detector pairs in SGWB detection~\cite{PhysRevD.59.102001}.

\subsection{The non-correlated configuration} \label{The non-correlated config}
We can also determine the sensitivity in the case where the signals in two detectors are non-correlated, i.e. the cross-correlation is negligible.
This condition is satisfied when either two detectors are well-separated (i.e. $|\Delta\mathbf{x}_{ab}|\gg\lambda_c$) or their sensitive axes are orthogonal.
In this case, the off-diagonal terms in $\Sigma$ vanish,
\begin{equation} \label{signal+noise corl matrix of noncorl case}
    \Sigma = N(f)\begin{bmatrix}
        1 & \\
          & 1
    \end{bmatrix}
    +\lambda^2\Gamma_X(f) S_{\Phi}(f)\begin{bmatrix}
        1 & \\
         & 1
    \end{bmatrix}
    = \left(N+\lambda^2\Gamma_X S_{\Phi}\right) I_{2\times2},
\end{equation}
and the likelihood reduces to
\begin{equation} \label{noncorl network likelihood diag}
    \mathcal{L}_{\text{stoch}}\left(\mathbf{\tilde{d}}\Big|\lambda, S_{\Phi}, N_1, N_2\right) = 
     \frac{1}{\pi^2T^2\left(N+\lambda^2\Gamma_X S_{\Phi}\right)^2}
    \exp\left[-\frac{\mathbf{\tilde{d}}^{\dagger}\mathbf{\tilde{d}}}{T(N+\lambda^2\Gamma_X S_{\Phi})}\right].
\end{equation}
Solving Eq.~(\ref{ML estimator eq}), the MLE of $\lambda^2$ is given by
\begin{equation}
    \hat{\lambda}^2 
    = \frac{\mathbf{\tilde{d}}^{\dagger}\mathbf{\tilde{d}}-2TN}{2T\Gamma_X S_{\Phi}}
    = \frac{\Re[\tilde{d}_1]^2+\Im[\tilde{d}_1]^2+\Re[\tilde{d}_2]^2+\Im[\tilde{d}_2]^2-2TN}{2T\Gamma_X S_{\Phi}}.
\end{equation}
According to Eq.~(\ref{noncorl network likelihood diag}), $\Re[\tilde{d}_1],\Im[\tilde{d}_1],\Re[\tilde{d}_2],\Im[\tilde{d}_2]$ are independent Gaussian variables with zero mean and variance $T(N+\lambda^2\Gamma_X S_{\Phi})/2$.
Therefore, the PDF of $\hat{\lambda}^2$ is linked to the $\chi^2$ distribution with $4$ degrees of freedom and can be obtained through a variable transformation. The result is
\begin{equation} \label{2d noncorl lambda dis}
    \mathcal{L}_{\text{stoch}}\left(\hat{\lambda}^2\Big|\lambda, S_{\Phi}, N_1, N_2\right)
    = \frac{\Gamma_X S_{\Phi}}{\varsigma}
    \left(\frac{\Gamma_X S_{\Phi}}{\varsigma}\hat{\lambda}^2+\frac{N}{\varsigma}\right) 
    \exp\left[-\left(\frac{\Gamma_X S_{\Phi}}{\varsigma}\hat{\lambda}^2+\frac{N}{\varsigma}\right)\right],
\end{equation}
where $\varsigma \equiv (N+\lambda^2\Gamma_X S_{\Phi})/2$.
Substituting Eq.~(\ref{2d noncorl lambda dis}) into Eqs.~(\ref{alpha+gamma1}, \ref{alpha+gamma2}), we obtain the sensitivity of the non-correlated configuration
\begin{equation} \label{lambda min non-corl}
    \lambda_{\min}^2 
    = \frac{N}{\Gamma_X S_{\Phi}} \left(\frac{x_{\alpha}}{x_{\gamma}}-1\right)
    = \frac{G(\alpha,\gamma)N}{2\Gamma_X S_{\Phi}} \left(\frac{\ln\alpha}{\ln\gamma}-1\right),
\end{equation}
where $x_z$ is determined by 
\begin{equation}
    z = \int^{\infty}_{x_z} dx\;p_4(x)
\end{equation}
with $p_4(x)$ the PDF of $\chi^2$ distribution with $4$ degrees of freedom.
The ratio $G(\alpha, \gamma) = 2(x_{\alpha}/x_{\gamma}-1)/(\ln\alpha/\ln\gamma-1)$ quantifies the relative sensitivity between the correlated and non-correlated configurations.
For $(\alpha, \gamma)=(0.05,0.95)$, $G\approx0.43$.
This implies that the sensitivity given by the non-correlated configuration is better. This contradicts the intuition derived from SGWB detection, where sensitivity generally benefits from correlation. However, as we will elucidate, this apparent contradiction can be resolved by considering the physical rationale behind preferring a non-correlated network in ULDM detection.

We go back to Eq.~$(\ref{1d power ratio})$, which indicates the relative strength between the ULDM signal and the detector noise. For $(\alpha, \gamma) = (0.05, 0.95)$, we observe that the minimum detectable value of $d_g$ corresponds to a power of signal that is $57$ times greater than that of noise. Even when using a two-detector network, which improves the sensitivity by a factor of $\mathcal{O}(1)$, the signal power remains notably higher than the noise power by an order of magnitude. 
In summary, we are searching for a relatively strong signal within the data, and detector noise does not pose a serious obstacle to signal detection. 
This differs from SGWB detection, where a weak signal is buried deeply in noise and correlation boosts sensitivity significantly.

While utilizing correlation to suppressing noise is not so substantial, as evidenced by the mere $\sqrt{2}$ enhancement, there is a significant advantage for ULDM detection when detectors are non-correlated.
In the co-aligned and co-located configuration, detectors sample a single patch of the field. This setup can lead to weak signals simultaneously if the amplitude of the field in the patch is close to zero due to fluctuation, even if the coupling strength $d_g$ is not small. On the contrast, when detectors are separated by a distance greater than the coherence length of ULDM, they will probe two patches with uncorrelated amplitudes, as illustrated in Fig.~\ref{fig:Phi}. This reduces the likelihood that both detectors meet a small amplitude and have weak signals simultaneously. 
Similar reasoning also applies for the orthogonal setup. As indicated by Eq.~(\ref{2pp gradient}), for an isotropic halo, the three spatial components of the field gradient at the same spacetime point are uncorrelated.~\footnote{For a halo with an axisymmetric DM velocity distribution, as described in Eq.~(43), we can construct a Cartesian coordinate system where the $z$ axis aligns with the symmetry axis of the velocity distribution.
In this coordinate system, the three spatial components of the field gradient are uncorrelated.}
In other words, the three spatial components can be viewed as three independent patches. Thus, detectors with orthogonal sensitive axes also effectively probe uncorrelated patches.

Therefore, the non-correlated configuration has a lower chance of missing ULDM signal compared to the correlated configuration.
The sensitivity gains from the less chance of missing compete with the loss resulting from the absence of correlation. Since we hope to detect ULDM signal with a high probability under a low false alarm rate, which corresponds to a strong signal, ``the less chance of missing" is more crucial, and the non-correlated configuration is preferred. 
While we focus on the network composed of GW interferometers, we expect our results will be also valid for other ULDM detection networks working in the coherence limit.

After identifying the non-correlated configuration as the optimal choice in ULDM detection,
an immediate question is that whether it is possible to implement such a configuration considering the planned space-based GW detectors.
Given that the distance between the Earth and the Sun is about 500 light seconds, the coherence length of ULDM in the sensitive band is either longer than or comparable to the distance between two detectors orbiting around the Sun in the Earth's orbit. Therefore, it seems unlikely to achieve the well-separated configuration with LISA, Taiji and TianQin. 
As comparison, the orthogonal configuration is more feasible. 
Moreover, the orthogonal configuration can even be accomplished with one detector. Since the detector orbits around the Sun, the direction of the sensitive axis evolves throughout the year. Therefore, we can correlate the data observed at different times when the directions of the sensitive axis are orthogonal, and form the orthogonal configuration virtually.

\section{Conclusions} \label{Conclusions}
In this work, we investigate the sensitivity of space-based gravitational-wave detectors to the stochastic ultralight dark matter field and illustrate with the model in which test masses are coupled with the gradient of scalar ULDM field. 
We find that, even after accounting for the stochastic effect, future missions such as LISA and Taiji are projected to improve the current best limit from MICROSCOPE, up to a factor of 1.7 with a one-year observation and a factor of 3 with a four-year observation around the most sensitive mass range $\sim 10^{-17}~$eV. 
Additionally, GW experiments will probe a different direction in parameter space compared to equivalence principle tests, making them complementary.

Given the potential of joint observations, we further explore the sensitivity of a detector network by applying the general correlation likelihood to ULDM detection. We introduce the overlap reduction function for ULDM and 
discuss how its behavior is influenced by the separation and relative orientation between detectors. We find that the overlap reduction function almost vanishes when the sensitive axes of detectors are orthogonal.
We analytically derive the sensitivity of the network in typical configurations and investigate the optimal configuration for ULDM detection.
We find that the non-correlated configuration, where two detectors observe independent ULDM signals, outperforms the correlated setup (co-located and co-aligned). In the non-correlated configuration, detectors probe independent patches of the ULDM field, reducing the probability that all patches have negligible amplitudes and increasing the chance of detecting a signal. 
Our findings provide new insights for potential joint observations involving space-based detectors like LISA and Taiji, as well as other ULDM detection networks operating in the coherence limit.

\section*{acknowledgement}
We would like to thank the anonymous referees for their valuable comments and suggestions, which have helped us improve the contents.
YT would like to thank Huai-Ke Guo for helpful discussions. This work is supported by the National Key Research and Development Program of China (No.2021YFC2201901), the National Natural Science Foundation of China (NSFC) (No.~12347103) and the Fundamental Research Funds for the Central Universities.

\appendix
\section{Derivation of the two-point correlation function} \label{appendix 2p func}
The two-point correlation function of the field is defined as
\begin{equation}
\begin{split}
    \left\langle \Phi(x)\Phi(x^{\prime}) \right\rangle &= \frac{\rho}{m^2} \left\langle 
    \sum_{n,l} \sqrt{\Delta_{nl}} \beta_{nl} \cos\left[m\left(1+\frac{v_{n}^2}{2}\right) t - \boldsymbol{k}_{nl}\cdot\boldsymbol{x} + \theta_{nl}\right] \right.\\
    &\qquad\qquad \left. \times \sum_{p,q} \sqrt{\Delta_{pq}} \beta_{pq} \cos\left[m\left(1+\frac{v_{p}^2}{2}\right) t^{\prime} - \boldsymbol{k}_{pq}\cdot\boldsymbol{x}^{\prime} + \theta_{pq}\right]
    \right\rangle.
\end{split}
\end{equation}
We split the sum over $n$, $l$, $p$, $q$ into the diagonal part $(n,l)=(p,q)$ and the off-diagonal part $(n,l)\neq(p,q)$. For the diagonal part,
\begin{equation}
\begin{split}
    &\sum_{n,l} \Delta_{nl} \times \left\langle\beta^2_{nl}\right\rangle 
    \times 
    \left\langle 
    \cos\left[m\left(1+\frac{v_{n}^2}{2}\right) t - \boldsymbol{k}_{nl}\cdot\boldsymbol{x} + \theta_{nl}\right] \right.\\
    &\qquad\qquad\qquad\qquad\qquad 
    \left.\cos\left[m\left(1+\frac{v_{n}^2}{2}\right) t^{\prime} - \boldsymbol{k}_{nl}\cdot\boldsymbol{x}^{\prime} + \theta_{nl}\right]
    \right\rangle \\
    &= \sum_{n,l} \Delta_{nl} \times 2 \times 
    \frac{1}{2}\cos\left[m\left(1+\frac{v_{n}^2}{2}\right) \tau - \boldsymbol{k}_{nl}\cdot\boldsymbol{d} \right],
\end{split}
\end{equation}
where $\tau=t^{\prime}-t$ and $\boldsymbol{d}=\boldsymbol{x}^{\prime}-\boldsymbol{x}$, and we use $\left\langle\beta^2_{nl}\right\rangle=2$. 
For the off-diagonal part,
\begin{equation}
    \sum_{(n,l)\neq(p,q)} \sqrt{\Delta_{nl}\Delta_{pq}} \times \left\langle\beta_{nl}\right\rangle \left\langle\beta_{pq}\right\rangle
    \times 
    \left\langle 
    \cos\left[m\left(1+\frac{v_{n}^2}{2}\right) t - \boldsymbol{k}_{nl}\cdot\boldsymbol{x} + \theta_{nl}\right] 
    \right\rangle
    \times
    \left\langle\cdots\right\rangle.
\end{equation}
The average of a single cosine function gives zero, and the off-diagonal part does not contribute to the two-point correlation function. Collecting the above results, we have 
\begin{equation}
    \left\langle \Phi(x)\Phi(x^{\prime}) \right\rangle = \frac{\rho}{m^2}
    \sum_{n,l} \Delta_{nl} \times  
    \cos\left[m\left(1+\frac{v_{n}^2}{2}\right) \tau - \boldsymbol{k}_{nl}\cdot\boldsymbol{d} \right],
\end{equation}
or written in the continuous form
\begin{equation}
    \left\langle \Phi(x)\Phi(x^{\prime}) \right\rangle = \frac{\rho}{m^2}
    \int d^3v\;f(\boldsymbol{v})  
    \cos\left[m\left(1+\frac{v^2}{2}\right) \tau - \boldsymbol{k}\cdot\boldsymbol{d} \right].
\end{equation}

\section{Two random phase models} \label{Random phase model}
There are two popular random phase models~\footnote{Although both models appear in literatures, it seems that the ER model is more natural since it arise from the classical limit of quantum field theory.} in the community.
One is the model presented in sec.~\ref{Theoretical Framework}, in which both the amplitudes and phases of plane waves are random variables~\cite{PhysRevA.97.042506,PhysRevD.97.123006,nakatsuka2022stochastic}, while in the other only the phases are random and the amplitudes are fixed~\cite{ Centers:2019dyn,Hui:2021tkt,Hui_2021}.
We denote the former as the entire-random model~(ER) and the later as the semi-random model~(SR).
Here, we briefly describe the SR model and compare it with the ER model. 
As we will show, the two models predict different strengths of the gradient correlation, and
consequently, the derived detector sensitivities will differ by an $\mathcal{O}(1)$ factor, depending on the random phase model employed.

The expression of $\Phi(x)$ in the SR model is given by
\begin{equation} \label{QR Phi}
    \Phi(x) = A\sum_{n}\sum_{l} A_{nl} e^{i\left(\omega_{nl}t - \boldsymbol{k}_{nl}\cdot\boldsymbol{x} + \theta_{nl}\right)},
\end{equation}
where $A$ is an overall constant fixed by the local energy density of DM, and $A_{nl}=g(\boldsymbol{k}_{nl})$ are the normalized amplitudes with $g(\boldsymbol{k})$ the momentum distribution of DM particles. Notice that $A_{nl}$ are not random variables.

The two-point correlation function of the field in the SR model is given by
\begin{equation}
    \left\langle \Phi(x)\Phi(x^{\prime}) \right\rangle^{\text{SR}} = \frac{A^2}{2}
    \int d^3k\;g^2(\boldsymbol{k})  
    \cos\left[m\left(1+\frac{v^2}{2}\right) \tau - \boldsymbol{k}\cdot\boldsymbol{d} \right].
\end{equation}
Changing the integral variable from $\boldsymbol{k}$ to $\boldsymbol{v}$, we have
\begin{equation}
    \left\langle \Phi(x)\Phi(x^{\prime})\right\rangle^{\text{SR}} = \frac{A^2}{2m^3}
    \int d^3v\;f^2(\boldsymbol{v})  
    \cos\left[m\left(1+\frac{v^2}{2}\right) \tau - \boldsymbol{k}\cdot\boldsymbol{d} \right],
\end{equation}
which depends on the square of the velocity distribution, while the dependence in Eq.~(\ref{2pp}) is linear.
Assuming that $\Phi$ constitutes the entire local DM and $f(\boldsymbol{v})$ follows Eq.~(\ref{boosted Maxwell}), we have
\begin{equation} \label{QR normalize}
    \rho = m^2\left\langle\Phi^2(x)\right\rangle^{\text{SR}} 
    = \frac{A^2}{2m}\int d^3v\;f^2(\boldsymbol{v}) 
    = \frac{A^2}{16m\left(\sqrt{\pi}\sigma\right)^3}.
\end{equation}
The two-point correlation function of the gradient field is given by
\begin{equation}
    \left\langle\partial_i\Phi(x)\partial'_j\Phi(x^{\prime})\right\rangle^{\text{SR}} = \frac{A^2}{2m}
    \int d^3v\;f^2(\boldsymbol{v})v_iv_j  
    \cos\left[m\left(1+\frac{v^2}{2}\right) \tau - \boldsymbol{k}\cdot\boldsymbol{d} \right].
\end{equation}
For $\tau=\boldsymbol{d}=0$, we have
\begin{equation}
\left\langle\left(\partial_x\Phi\right)^2\right\rangle^{\text{SR}}
    = \left\langle\left(\partial_y\Phi\right)^2\right\rangle^{\text{SR}}
    = \frac{\rho\sigma^2}{2},\qquad
    \left\langle\left(\partial_z\Phi\right)^2\right\rangle^{\text{SR}}
    = \frac{3\rho\sigma^2}{2},
\end{equation}
where we set $\boldsymbol{v}_{\odot}=\sigma\hat{z}$ in Eq.~(\ref{boosted Maxwell}). 
As comparison, in the ER model presented in sec.~\ref{Theoretical Framework}, we have
\begin{equation}
    \left\langle\left(\partial_x\Phi\right)^2\right\rangle
    = \left\langle\left(\partial_y\Phi\right)^2\right\rangle
    = \rho\sigma^2,\qquad
    \left\langle\left(\partial_z\Phi\right)^2\right\rangle
    = 2\rho\sigma^2.
\end{equation}
Therefore, even for the same velocity distribution, there is an $\mathcal{O}(1)$ difference in the 
strengths of the gradient correlation reported by the SR and ER models, despite both models normalizing the two-point correlation of the field to the local DM density.

Besides the choice of random phase model, discrepancies of values such as $d_g^{\text{stoch}}/d_g^{\text{det}}$ between the literature can arise from the differences in the value of DM velocity dispersion and the velocity of the laboratory in the galactic-rest frame used.
In the supplementary material of Ref.~\cite{Centers:2019dyn}, the authors adopted the SR model and assumed that
\begin{equation}
    A_{\boldsymbol{k}} \propto \exp\left[-\frac{k^2_x+k^2_y+(k_z-k_{\text{lab}})^2}{k^2_0}\right],
    \qquad k_{\text{lab}} \simeq k_0,
\end{equation}
which in our notation corresponds to a higher laboratory velocity $\boldsymbol{v}_{\odot}=\sqrt{2}\sigma\hat{z}$, and results in that $\left\langle\left(\partial_x\Phi\right)^2\right\rangle^{\text{SR}}_{\text{Centers}}=\left\langle\left(\partial_y\Phi\right)^2\right\rangle^{\text{SR}}_{\text{Centers}}=2\rho\sigma^2$ and
$\left\langle\left(\partial_z\Phi\right)^2\right\rangle^{\text{SR}}_{\text{Centers}}=5\rho\sigma^2/2$.
Additionally, they assumed $\left(\partial_i\Phi^{\text{det}}\right)^2_{\text{Centers}}\simeq2\rho\sigma^2$ in the deterministic case, while we adopt $\left(\partial_i\Phi^{\text{det}}\right)^2\simeq\rho\sigma^2/3$ to account for the randomness in the direction of the field gradient.
Substituting the above numerical values into Eq.~(\ref{C ratio}), we broadly recover their result
\begin{equation} \label{ratio estimation}
    \frac{d_g^{\text{stoch}}}{d_g^{\text{det}}}
    \simeq \sqrt{\frac{\left(\partial_i\Phi^{\text{det}}\right)^2_{\text{Centers}}}{\left\langle(\partial_z\Phi)^2\right\rangle^{\text{SR}}_{\text{Centers}}}C} \approx 2.4.
\end{equation}

\section{The power spectral density of noise} \label{noise psd}
The one-sided power spectral density of the noise in $X$ channel is given by
\begin{equation}
    N_{X} = 16\sin^{2}(2\pi fL) 
    \left\{\left[ 3+\cos(4\pi fL) \right]S_{\text{acc}} + S_{\text{oms}}\right\},
\end{equation}
where $L$ is the armlength of interferometer, and $S_{\text{oms}}$ and $S_{\text{acc}}$ stand for the optical metrology system noise and test mass acceleration noise, respectively. They are further described as 
\begin{align}
    \label{oms} S_{\text{oms}}\left( f\right) &= \left(\frac{2\pi f}{c}s_{\text{oms}}\right)^{2} \left[1 + \left(\frac{2 \times 10^{-3}~\textrm{Hz}}{f} \right)^{4}\right]
    \;\frac{1}{\textrm{Hz}}, \\
    \label{acc} S_{\text{acc}}\left( f\right) &= \left( \frac{s_{\text{acc}}}{2\pi fc} \right)^{2} \left[ 1+\left(\frac{0.4 \times 10^{-3}~\textrm{Hz}}{f} \right)^{2} \right] \; \left[ 1+ \left(\frac{f}{8 \times 10^{-3}~\textrm{Hz}} \right)^{4} \right]
    \;\frac{1}{\textrm{Hz}},
\end{align}
where we adopt the following parameters for LISA~\cite{amaroseoane2017laser}, Taiji~\cite{Hu:2017mde} and TianQin~\cite{Luo_2016}:
\begin{align*}
   \textrm{LISA} : s_{\text{oms}} &= 15\times10^{-12}~\textrm{m},
   \;& s_{\text{acc}} &= 3\times10^{-15}~\textrm{m}/\textrm{s}^2,
   \;& L &= 2.5\times10^6~\textrm{km}; \nonumber \\
   \textrm{Taiji} : s_{\text{oms}} &= 8\times10^{-12}~\textrm{m},
   \;& s_{\text{acc}} &= 3\times10^{-15}~\textrm{m}/\textrm{s}^2,
   \;& L &= 3\times10^6~\textrm{km}; \nonumber \\
   \textrm{TianQin} : s_{\text{oms}} &= 1\times10^{-12}~\textrm{m},
   \;& s_{\text{acc}} &= 1\times10^{-15}~\textrm{m}/\textrm{s}^2,
   \;& L &= 1.7\times10^5~\textrm{km}.
\end{align*}

\section{Estimation of the sensitivity of the LISA-Taiji network}
In reality, LISA and Taiji will have different noise performances and arm lengths. In this appendix, we account for these differences and estimate the sensitivities of the LISA-Taiji network under the optimal configurations for the ULDM detection.

The MLE is still determined by Eq.~(\ref{ML estimator eq}) but now with 
\begin{equation} \label{LISA-Taiji matrix}
    \Sigma = \begin{bmatrix}
        N_1 & \\
          & N_2
    \end{bmatrix}
    + \lambda^2\begin{bmatrix}
        \Gamma_1 S_{\Phi} & \\ 
         & \Gamma_2 S_{\Phi}
    \end{bmatrix},
\end{equation}
where $N_i$s are the noise PSD of LISA and Taiji, respectively, and $\Gamma_i$s are the square of modulus of the $X$ channel response function. Substituting Eq.~(\ref{LISA-Taiji matrix}) into Eq.~(\ref{ML estimator eq}), we have
\begin{equation}
    \frac{\Gamma_1S_{\Phi}}{N_1+\hat{\lambda}^2\Gamma_1S_{\Phi}}\left[1-\frac{|\tilde{d}_1|^2}{T(N_1+\hat{\lambda}^2\Gamma_1S_{\Phi})}\right]
    + \frac{\Gamma_2S_{\Phi}}{N_2+\hat{\lambda}^2\Gamma_2S_{\Phi}}\left[1-\frac{|\tilde{d}_2|^2}{T(N_2+\hat{\lambda}^2\Gamma_2S_{\Phi})}\right] = 0.
\end{equation}
In the strong signal limit, the approximate solution of the above equation is 
\begin{equation}
    \hat{\lambda}^2 \simeq \frac{|\tilde{d}_1|^2-TN_1}{2T\Gamma_1S_{\Phi}} + \frac{|\tilde{d}_2|^2-TN_2}{2T\Gamma_2S_{\Phi}}.
\end{equation}
We define a new set of random variables 
\begin{equation}
    \begin{aligned}
        u_1 &= \sqrt{\frac{2}{T(N_1+\lambda^2\Gamma_1S_{\Phi})}}\Re[\tilde{d}_1],\qquad 
        v_1 = \sqrt{\frac{2}{T(N_1+\lambda^2\Gamma_1S_{\Phi})}}\Im[\tilde{d}_1], \\
        u_2 &= \sqrt{\frac{2}{T(N_2+\lambda^2\Gamma_2S_{\Phi})}}\Re[\tilde{d}_2],\qquad 
        v_2 = \sqrt{\frac{2}{T(N_2+\lambda^2\Gamma_2S_{\Phi})}}\Im[\tilde{d}_2], \\
    \end{aligned}
\end{equation}
under which the likelihood of data is again given by Eq.~(\ref{optimal network likelihood diag}), and 
\begin{equation}
\begin{split}
    \hat{\lambda}^2 
    &= \frac{\lambda^2\Gamma_1S_{\Phi}+N_1}{\Gamma_1S_{\Phi}}(u^2_1+v^2_1)
    + \frac{\lambda^2\Gamma_2S_{\Phi}+N_2}{\Gamma_2S_{\Phi}}(u^2_2+v^2_2)
    - \frac{1}{2}\left(\frac{N_1}{\Gamma_1S_{\Phi}}+\frac{N_2}{\Gamma_2S_{\Phi}}\right) \\
    &\simeq \left(\lambda^2+\overline{\frac{N}{\Gamma S_{\Phi}}}\right)\frac{u^2_1+v^2_1+u^2_2+v^2_2}{4} -\overline{\frac{N}{\Gamma S_{\Phi}}},
\end{split}
\end{equation}
where we define
\begin{equation}
    \overline{\frac{N}{\Gamma S_{\Phi}}} = \frac{1}{2}\left(\frac{N_1}{\Gamma_1S_{\Phi}}+\frac{N_2}{\Gamma_2S_{\Phi}}\right).
\end{equation}
Following exactly the same statistical approach outlined in~\ref{Network}, the sensitivity of the LISA-Taiji network in the optimal configuration is approximately given by
\begin{equation}
    \lambda^2_{\min} = \overline{\frac{N}{\Gamma S_{\Phi}}}
    \left(\frac{x_{\alpha}}{x_{\gamma}}-1\right),
\end{equation}
which has the same form as Eq.~(\ref{lambda min non-corl}) but with $N/\Gamma S_{\Phi}$ replaced by the average quantity of LISA and Taiji.

\section{Derivation of the ``sky map" of ULDM} \label{appendix DM skymap}
Using the plane-wave expansion, we write $\Phi(x)$ as
\begin{equation} \label{phi plane-wave expansion}
    \Phi(x) = \int^{\infty}_{-\infty} df \int d\hat{k} \;\Phi(f,\hat{k})e^{i2\pi f\left(t-\eta\hat{k}\cdot\mathbf{x}\right)} ,
\end{equation}
where $\eta=\sqrt{1-f^2_c/f^2}$ 
and the integral $\int^{\infty}_{-\infty}df$ should be understood as $\int^{\infty}_{f_c}df+\int^{-f_c}_{-\infty}df$.
Since $\Phi(x)$ is real, we have $\Phi(f,\hat{k})=\Phi^*(-f,\hat{k})$.
The two-sided power spectral density $S_{\Phi}(f,\hat{k})$ is defined as
\begin{equation} \label{DM skymap}
    \left\langle \Phi^*(f,\hat{k})\Phi(f^{\prime},\hat{k}^{\prime})\right\rangle
    = \delta\left(f-f^{\prime}\right) \; \delta^2\left(\hat{k}, \hat{k}^{\prime}\right) \times S_{\Phi}(f,\hat{k}),
\end{equation}
which quantifies the power of ULDM concentrated in the infinitesimal interval $dfd\hat{k}$ centered at frequency $f$ and sky direction $\hat{k}$.

We can recalculate the two-point correlation function using Eqs.~(\ref{phi plane-wave expansion}, \ref{DM skymap})
\begin{equation} \label{corl skymap}
\begin{aligned}
    \left\langle\Phi(x)\Phi(x')\right\rangle &= 
    \left\langle \int^{\infty}_{-\infty} df \int d\hat{k}\;\Phi^*(f,\hat{k})e^{-i2\pi f\left(t-\eta\hat{k}\cdot\boldsymbol{x}\right)} \right.\\
    &\qquad\qquad\qquad\qquad\qquad\qquad
    \left.\times\int^{\infty}_{-\infty} df^{\prime} \int d\hat{k}^{\prime}\;\Phi(f^{\prime},\hat{k}^{\prime})e^{i2\pi f^{\prime}\left(t^{\prime}-\eta^{\prime}\hat{k}^{\prime}\cdot\boldsymbol{x}^{\prime}\right)} \right\rangle \\
    &= \int df df^{\prime} \int d\hat{k} d\hat{k}^{\prime} \;
    e^{-i2\pi f\left(t-\eta\hat{k}\cdot\boldsymbol{x}\right)} e^{i2\pi f^{\prime}\left(t^{\prime}-\eta^{\prime}\hat{k}^{\prime}\cdot\boldsymbol{x}^{\prime}\right)} \\
    &\qquad\qquad\qquad\qquad\qquad\qquad
    \times\delta\left(f-f^{\prime}\right) \; \delta^2\left(\hat{k}, \hat{k}^{\prime}\right) \times S_{\Phi}(f,\hat{k}) \\
    &= \int^{\infty}_{-\infty} df \int d\hat{k}\;
    e^{i2\pi f\left(\tau-\eta\hat{k}\cdot\boldsymbol{d}\right)} \times S_{\Phi}(f,\hat{k}) \\
    &= 2\int^{\infty}_{f_c} df \int d\hat{k}\;
    \cos\left(2\pi f\tau-\boldsymbol{k}\cdot\boldsymbol{d}\right) \times S_{\Phi}(f,\hat{k})\;,   
\end{aligned}
\end{equation}
where we use $S_{\Phi}(f,\hat{k})=S_{\Phi}(-f,\hat{k})$ in the last line.
The two-point correlation derived in Appendix~\ref{appendix 2p func} can be recast into
\begin{equation} \label{corl velocity}
    \begin{aligned}
    \left\langle\Phi(x)\Phi(x')\right\rangle  
    &= \frac{\rho}{m^2}
    \int \frac{d^3k}{m^3}\;
    \cos\left(2\pi f\tau - \boldsymbol{k}\cdot\boldsymbol{d}\right)
    f\left(\frac{\boldsymbol{k}}{m}\right)\\
    &= \int^{\infty}_{f_c}df \int d\hat{k}\; 
    \cos\left(2\pi f\tau-\boldsymbol{k}\cdot\boldsymbol{d}\right) \times
    \frac{2\pi \rho v_f}{m^3}f(v_f\hat{k})\;,
    \end{aligned}
\end{equation}
where $v_f=\sqrt{2(f/f_c-1)}$, and we use the non-relativistic approximation $2\pi f\simeq m(1+v^2/2)$ and $\boldsymbol{k}=m\boldsymbol{v}$. Comparing Eq.~(\ref{corl skymap}) with Eq.~(\ref{corl velocity}), we have 
\begin{equation}
    S_{\Phi}(f,\hat{k}) = \frac{\pi\rho v_f}{m^3} f(v_f\hat{k}).
\end{equation}

\bibliography{ref}
\end{document}